\documentclass[journal]{IEEEtran}
\usepackage{algorithmic}
\usepackage{amsmath,amsfonts}
\usepackage{algorithm}
\usepackage{array}
\usepackage[caption=false,font=normalsize,labelfont=sf,textfont=sf]{subfig}
\usepackage{textcomp}
\usepackage{stfloats}
\usepackage{url}
\usepackage{verbatim}
\usepackage{graphicx}
\usepackage{multirow}
\usepackage{booktabs}

\newcommand{\bestinblock}[1]{\textbf{#1}}          % Best in block (bold)
\newcommand{\bestoverall}[1]{\underline{\bestinblock{#1}$^\star$}} % Best overall with star

\newcommand{\amfontsize}[1]{#1}          % Small caps for acronyms

\newcommand{\X}{\mathbf{X}}                        % bold math X
\newcommand{\W}{{W}}                               % W, can be changed to \mathbf{W} if preferred
\newcommand{\Y}{{Y}}                               % Y
\newcommand{\A}{\mathcal{A}}                       % calligraphic A
\newcommand{\x}{\mathbf{x}}                        % bold math x
\newcommand{\R}{\mathbb{R}}                        % Real numbers symbol
\newcommand{\acronym}[1]{\textbf{\underline{#1}}} % For acronyms with underline + bold

\begin{document}

\title{Large Language Model Guided Decoding for Self-Supervised Speech Recognition}

% Large Language Model Guided Decoding for Self-Supervised Speech Recognition

% Large Language Model Augmented Decoding for Self-Supervised Speech Recognition

% Large Language Model Driven Decoding for Self-Supervised Speech Recognition

\author{Eyal Cohen, Bhiksha Raj,~\IEEEmembership{Fellow,~IEEE,}% 
        and Joseph Keshet,~\IEEEmembership{Senior~Member,~IEEE}%
\thanks{E. Cohen and J. Keshet are with the Faculty of Electrical and Computer Engineering, Technion-Israel Institute of Technology (e-mail: eyalcohen308@gmail.com; jkeshet@technion.ac.il).}%
\thanks{B. Raj is with Carnegie Mellon University, Pittsburgh, PA, USA (e-mail: bhiksha@cs.cmu.edu).}%
}

% The paper headers
% \markboth{IEEE Journal of Selected Topics in Signal Processing}%
%         {Cohen \MakeLowercase{\textit{et al.}}: Guiding an ASR Decoder using LLMs}

\maketitle

\begin{abstract}
Self-supervised automatic speech recognition (SSL-ASR) is an ASR approach that uses speech encoders pretrained on large amounts of unlabeled audio (e.g., wav2vec2.0 or HuBERT) and then fine-tunes them with limited labeled data to perform transcription. Decoding is usually performed with a CTC decoder, whose hypotheses are scored and refined using an external language model (LM), typically an n-gram or neural LM, which guides beam search to produce the final transcription. Using Large Language Models (LLMs) as external LMs remains a challenge, as their word probabilities are overly confident. 

The proposed method integrates an LLM with an SSL acoustic model by using the LLM's decoding mechanism to generate a set of candidate next tokens. For each candidate, the SSL model provides an acoustic score by aligning it to the input acoustics of the SSL model. A combined acoustic and LLM score is then calculated based on decomposing the MAP estimator of words given the acoustic signal. The tokens with the highest combined scores are maintained in a beam, which is then used to proceed to the next decoding step. We illustrate the effectiveness of our method through a comprehensive comparison with the current state-of-the-art LLM-based decoding, post-processing, and error-correcting methods across multiple datasets. Our approach proves particularly effective when processing challenging inputs such as complex speech sentences, acronyms, and domain-specific vocabulary. 
%our method allows the llm to be improved independently and that can directly reflect on asr output without any retraining
\end{abstract}

% File: introduction.tex
\section{Introduction}
\label{sec:intro}

% Significance of ASR in Modern Technology
%\IEEEPARstart{A}{utomatic} Speech Recognition (ASR) has witnessed significant advancements in recent years, driven by the growing capabilities of deep learning and large language models. ASR stands as a cornerstone of intelligent speech technology, enabling machines to understand and transcribe human speech with remarkable precision. It has become an integral part of modern digital experiences, powering applications like virtual assistants, real-time captions on social media platforms, and transcription services for podcasts and meetings. %As ASR technology approaches human-like accuracy, it is poised to enhance accessibility to digital content further and drive innovation across numerous fields.

Self-supervised automatic speech recognition (SSL-ASR) refers to ASR systems that use speech encoders pretrained on unlabeled audio using self-supervised learning methods such as wav2vec 2.0 \cite{baevski2020wav2vec}, HuBERT \cite{hsu2021hubert}, or WavLM \cite{chen2022wavlm}. These pretrained encoders are then fine-tuned with a small amount of labeled data---typically using CTC or a light decoder---to perform speech recognition. SSL-ASR is especially effective in low-resource or domain-adaptation settings because it decouples representation learning from the need for large supervised datasets. 
Although supervised and weakly supervised approaches such as Whisper \cite{radford2023whisper} have been shown to approach human-level transcription accuracy \cite{kim2024automatic}, there remain scenarios, particularly in low-resource conditions, domain adaptation, or child speech, where SSL-based ASR models are more suitable and often preferred \cite{barcovschi2023comparative, anidjar2024whisper, rouditchenko2023comparison, adnan2025one}.

% Modern ASR Architecture: AM, LM, and Decoding
SSL-ASR systems generally consist of three core components: the Acoustic Model (AM), the Language Model (LM), and the decoding mechanism. The AM converts raw audio signals into probabilistic representations of linguistic units, such as phones or sub-word units, capturing the acoustic features necessary for transcription. In this paper, we restrict ourselves to SSL-based models. The LM component ensures that the generated sequences are linguistically coherent by predicting the probability of word sequences based on extensive text corpora. The decoding mechanism integrates outputs from both components, utilizing algorithms such as beam search to identify the most likely transcription. %Together, these components form the backbone of SSL-ASR systems, enabling accurate and contextually relevant transcription of spoken language.

The LM component has traditionally relied on N-gram models, which are probabilistic models predicting the likelihood of a word sequence based on the preceding N-1 words \cite{jm3}. These models derive their probabilities from extensive text corpora, employing Markov assumptions to simplify calculations by limiting context. Although N-gram models are efficient and interpretable, they face challenges with data sparsity and long-range dependencies, rendering them less effective compared to modern neural language models in ASR. 

Advances beyond traditional N-gram models have led to substantial improvements in language modeling, driven largely by the adoption of more expressive deep learning architectures. One notable approach is the Gated CNN (GCNN)~\cite{dauphin2017language}, which uses stacked causal convolutions to create a neural language model that effectively captures local context within fixed-length token windows during decoding. Another significant model is the TransformerLM~\cite{baevski2018adaptive}, which employs multi-head self-attention to model global dependencies, allowing it to condition on the entire history of hypotheses. These developments mark a clear transition toward more powerful and flexible language modeling paradigms that overcome key limitations of conventional N-gram approaches.

% Large Language Models: Transforming NLP and Beyond
%Large Language Models (LLMs) have redefined natural language processing (NLP), achieving remarkable success in tasks like machine translation, conversational AI, and text summarization \cite{radford2019language, touvron2023llama}. Pre-trained on diverse textual corpora, LLMs excel at capturing linguistic nuances, contextual relationships, and semantic meanings. Their ability to generalize across tasks has positioned them as foundational tools in modern AI research, offering significant potential for applications beyond text, including speech processing.

Large Language Models (LLMs) have achieved remarkable success in various tasks, including machine translation, conversational AI, and text summarization \cite{radford2019language, touvron2023llama}. Pre-trained on diverse textual corpora, LLMs excel at capturing linguistic nuances, contextual relationships, and semantic meanings. %As ASR technology approaches human-like accuracy \cite{kim2024automatic}, 
It is only natural to try to replace traditional LMs with LLMs and leverage their advanced contextual understanding to further enhance recognition quality and downstream applications.

% Integrating LLMs into ASR
The integration of LLMs into SSL-ASR systems introduces some challenges \cite{toshniwal2018comparison}. Unlike traditional LMs, LLMs provide a deeper understanding of language and context, and can enable SSL-ASR systems to handle ambiguities in acoustic signals and domain-specific vocabulary more accurately. Although numerous efforts have been made to incorporate LLMs into SSL-ASR systems \cite{toshniwal2018comparison, ma2023can, tur2024progres}, these approaches typically focus on prompt design, output re-scoring, error-correction, or architectural modifications. A comprehensive review of these methods is provided in Section~\ref{sec:related_work}. Notably, however, none of these approaches integrates the LLM with the AM in a manner that enables mathematically principled inference through direct maximization of the posterior probability (MAP) of the word sequence given the acoustic input. It is worth noting that Spoken Language Models (SLMs) are constructed with components that are functionally analogous to both AMs and LLMs. However, their primary goal is not necessarily ASR; instead, they are designed to function as universal speech processing systems, capable of supporting a wide range of downstream tasks \cite{arora2025landscape}.
%Current LLM-based ASR architectures often combine three primary components: a speech encoder, a modality alignment projector, and the LLM itself \cite{wu2023decoder, tang2024alignment}. The encoder processes raw audio into feature representations, the projector aligns these outputs with the LLM's input space, and the LLM generates transcriptions with a holistic understanding of language. 
% Challenges in LLM-ASR Integration
% 

Our goal is to propose a method for integrating LLMs into SSL-based acoustic models while keeping them separable, allowing each to be independently trained and improved on its own data, and enabling the system to benefit from the strengths of both components without requiring joint optimization.
% [29/06/2025, 19:51:37] Bhiksha Raj: our pitch is that we want to be able to keep the acoustic model and lm separate, so that they can both be independently improved on their own data,  and only use the latter to guide the former to get the best of both worlds.

We propose a new decoding mechanism guided by the LLM, formally derived from a decomposition of the MAP estimator for the word sequence given the speech signal. The decoding process is iterative. In each iteration, the LLM samples a set of candidate tokens conditioned on the previously predicted tokens. The candidate tokens are aligned with the speech signal using the SSL model, which outputs an acoustic score for each alignment. The candidates with the highest combined score from both the SSL model and the LLM are added to the beam and incorporated into the prompt for the next iteration. The process concludes when all beams reach the end of the sentence or when a predetermined safety horizon is met. 

%%% We introduce a new iterative, LLM-guided decoder derived from the MAP estimate of the word sequence given speech. In each step, the LLM proposes candidate tokens from prior outputs; an SSL model aligns them to the signal and assigns acoustic scores. Tokens with the best combined LLM+SSL score extend the beam and update the prompt. Decoding stops when beams finish or a safety horizon is reached.

Our method operates in a zero-shot manner, meaning no additional training is required. This approach enables enhancements to the SSL model or LLM to be implemented independently, directly affecting ASR output without retraining. However, a notable drawback of our approach is its higher computational demands compared to standard inference, which we will discuss in further detail.

We demonstrate the effectiveness of the proposed method across three speech corpora, each showcasing a distinct speaking style: ALLSTTAR, which features short, simple sentences; WSJ, comprising read speech from newspapers; and TED-LIUM, consisting of presentations by real-world professionals. In the experiments, we used two state-of-the-art acoustic models (wav2vec 2.0, HuBERT) and three LLMs (GPT-2, LLaMA 2, Falcon). The integration of LLM into ASR proves advantageous for more complex speaking styles, enhancing its ability to manage acronyms and specialized vocabulary. Lastly, we compare out method with other decoding techniques, post-processing methods and LLM-based error-correcting methods. Our method shows particular strength in handling complex speech nuances such as acronyms and domain-specific vocabulary

% [29/06/2025, 19:52:55] Bhiksha Raj: the fact that our method allows the llm to be improved independently and that can directly reflect on asr output without any retraining.  Ditto for the AM.  This needs some highlighting

The remainder of this paper is structured as follows. Section~\ref{sec:related_work} reviews prior work, with an emphasis on integrating LLMs into ASR systems. In Section~\ref{sec:background}, we define the problem, introduce the notation, and provide an overview of conventional ASR decoding. Our proposed approach and its efficient solution, including pseudo-code, are detailed in Section~\ref{sec:model}. Section~\ref{sec:datasets} describes the datasets used for evaluation, while Section~\ref{sec:experiments} presents experimental results and comparisons with baseline methods. Finally, we conclude in Section~\ref{sec:conclusions}.
% Our work introduction
% In this work, we present a novel framework for enhancing the decoding process in ASR systems by leveraging the power of LLMs as the central driver of the decoding mechanism. Unlike previous approaches that require retraining or fine-tuning of speech encoders and projectors, our method directly utilizes pre-trained LLMs, LLaMA 2 \cite{llama2}, Falcon \cite{falcon}, alongside established acoustic models like wav2vec 2.0 and HuBERT, without additional training. This zero-shot approach enables seamless integration of LLMs to dynamically guide decoding, ensuring better alignment with acoustic inputs and delivering heightened transcription accuracy.

\section{Related Work}
\label{sec:related_work}

In traditional ASR systems, decoding is commonly formulated as a MAP search that integrates acoustic and language model scores. In the context of SSL-CTC acoustic models, this is most commonly realized through \emph{shallow fusion} with external language models, including traditional N-gram models \cite{jm3} as well as neural approaches such as Gated CNNs \cite{dauphin2017language} and Transformer-based language models \cite{baevski2018adaptive}.

The main line of work applies LLMs in a post-hoc manner, operating on the final ASR transcription to refine and improve the output. One such approach is \emph{N-best rescoring}, where a language model assigns scores to multiple candidate hypotheses produced by the ASR system and selects the most probable one \cite{chelba2012large, xu2018pruned, udagawa2022effect, xu2022rescorebert}. Another post-hoc approach is \emph{error correction}, in which language models directly correct or refine the ASR output. \cite{ma2023can} showed that general-purpose LLMs can perform ASR error correction in a zero-shot or one-shot setting by prompting the model with the ASR N-best hypotheses. They also demonstrated a supervised variant based on a fine-tuned T5 model operating on the same N-best input. \cite{ma2024asr} extended this work with additional post-hoc correction strategies, including constrained variants that operate over N-best lists or lattices. \cite{tur2024progres} proposed a hybrid post-hoc approach that augments the ASR N-best list with LLM-generated hypotheses and then jointly rescores both original and generated candidates using additional ASR confidence scores and LLM sequence scoring.

A complementary approach integrates LLMs into ASR systems by modifying the underlying architecture or through joint training of speech-text models. Some approaches introduce trainable projection layers that connect SSL encoder representations to an LLM \cite{ma2024embarrassingly, mundnich2024zero}. Other methods train multimodal speech-language models used for multiple downstream tasks \cite{deng2024transducer, jia2024efficient}. Although effective, these systems require supervised training or architectural changes and are therefore out of scope for this work.

\section{Background}
\label{sec:background}

The speech recognition task involves mapping an acoustic signal to its corresponding textual transcription. Formally, let $\X = (\x_1, \x_2, \dots, \x_T)$ represent the input audio sequence of length $T$, and $\x_t \in \R^d$ is a frame of speech, where $t \in [1, T]$ is the frame index. Let $\W = (w_1, w_2, \dots, w_K)$ denote the corresponding spoken word sequence. We assume that $K$ words were spoken, and $w_k \in \mathcal{V}$ for $1\le k \le K$ and $\mathcal{V}$ is the vocabulary. The objective is to determine the most probable transcription $\W^*$ given the acoustic input $\X$. In other words, the recognizer should maximize the probability $P(\W|\X)$ of the word sequence $\W$ spoken within $\X$ \cite{Jelinek1998}:
\begin{equation}\label{eq:map1}
\W^* = \arg\max_{\W} P(\W | \X).
\end{equation}

Traditionally, this MAP estimator is not used directly; instead, it is decomposed into two parts using Bayes' rule as follows:   
\begin{equation}\label{eq:map}
    P(\W | \X) = \frac{P(\W)P(\X|\W)}{P(\X)}~.
\end{equation}
Since the probability of $P(\X)$ does not change the optimal word sequence we can write \eqref{eq:map} as
\begin{equation}
    \W^* = \arg\max_{\W} P(\X|\W)P(\W)~,
\end{equation}
where $P(\X|\W)$ represents the probability of the audio sequence given the word sequence, and it is referred to as the \emph{acoustic model}, and the probability $P(\W)$ represents the probability of the word sequence and is referred to as the \emph{language model}. The process of identifying the word sequence $\W^*$ that maximizes the MAP estimate is known as \emph{decoding}. We provide a brief overview of each component, focusing on aspects pertinent to our proposed approach.

\subsection{Acoustic model} 

Traditionally, the acoustic model $P(\X | \W)$ was estimated using HMMs. In its standard form, an individual HMM is defined for each phone, and word-level HMMs are constructed by sequentially concatenating the phone-specific models. To represent an entire utterance, the HMMs corresponding to each word in the sequence $\W$ are further concatenated, yielding a single HMM that models the full spoken input \cite{rabiner1993fundamentals}. The model presumes that the probability of transitioning to a particular state depends solely on the current state, independent of the previous states. Additionally, the output probability of a speech frame is considered conditionally independent of prior or future frames, given the current state \cite{Jelinek1998}. HMM-based acoustic models lag behind modern transformer-based models. 

Our work focuses on acoustic modeling approaches based on self-supervised representations, including wav2vec 2.0 \cite{baevski2020wav2vec} and HuBERT \cite{hsu2021hubert}. These models are initially trained on raw, unlabeled audio to learn contextualized representations of speech. Then, they are fine-tuned on transcribed speech data using the Connectionist Temporal Classification (CTC) loss function \cite{graves2006connectionist}, enabling them to serve as effective acoustic models for automatic speech recognition. This loss enables them to operate directly on character sequences, rather than phonetic states, without requiring explicit alignment between the input audio and target output. As a result, these models adopt a fundamentally different formulation of the acoustic modeling task. Instead of estimating $P(\X|\W)$, they estimate the posterior probabilities $P(\Y|\X)$, where $\Y$ represents a sequence of $U$ characters, including the blank token $\epsilon$. Specifically, $\Y=(y_1, \ldots, y_U)$, with $y_i \in \{a,\ldots, z,\epsilon\}$ \cite{baevski2020wav2vec, hsu2021hubert}.

\subsection{Language model}

The language model plays a critical role in ensuring that the recognized text is linguistically plausible and contextually coherent. Its primary goal is to assign a probability to a given sequence of words and guide the decoding process toward the most likely interpretation of the spoken input. The language model probability $P(\W)$ is often decomposed as
\begin{equation}\label{eq:lm_cond}
    P(\W) = \prod_{n=1}^{N} P(w_n | w_1^{n-1})~,
\end{equation}
where $w_1^{n-1}=(w_1, \ldots, w_{n-1})$. The most commonly used language model used with SSL acoustic models is the N-gram language model \cite{baevski2020wav2vec, hsu2021hubert}, but GCNNs \cite{dauphin2017language}, and TransformerLM \cite{baevski2018adaptive} are used.

LLMs, such as GPT-2 \cite{radford2019language} and LLaMA 2 \cite{touvron2023llama}, are deep neural networks based on the transformer architecture, trained to predict the next token (typically a sub-word unit) given a sequence of preceding tokens. Although LLMs follow the same conditional formulation as in \eqref{eq:lm_cond}, they differ fundamentally from N-gram models. LLMs are trained using the cross-entropy loss over massive text corpora to learn the parameters of a transformer-based architecture. This training often results in a high norm of the output logit weights, which causes the softmax function to produce an overly sharp, overconfident probability distribution \cite{pmlr-v162-wei22d}.

Despite their strong generative capabilities, the effective use of LLMs as language models for SSL acoustic models remains challenging. While LLM models predict the probability of the next token given the sequence of previous tokens, the resulting probability is not calibrated due to the way they are trained \cite{Bridle1989}. Furthermore, ASR systems require language models that support incremental scoring, meaning they can evaluate partial word sequences (prefixes), manage multiple parallel hypotheses during beam search, and synchronize closely with the acoustic model's step-by-step output \cite{Jelinek1998}. However, LLMs are built to process complete sequences in a forward-only manner, making them ill-suited for this role. They lack native support for prefix scoring, cannot easily backtrack or compare multiple evolving hypotheses, and perform best when given the entire context, which is often unavailable in real-time ASR decoding. Since LLMs cannot be directly integrated into the acoustic model, we propose to incorporate them through the decoding process.

\subsection{Decoding}

The decoding mechanism refers to the process of inferring the most probable sequence of words or tokens given an input speech signal. It combines the outputs of the acoustic model and the language model to generate the final transcription using an efficient search algorithm, such as beam search, stack decoding, or fast match \cite{huang2001spoken}. Before introducing our decoding method, we briefly review decoding in HMM-based ASR and in SSL-ASR systems. This comparison provides context for evaluating our approach relative to the former, which is somewhat probabilistically principled (ignoring $\alpha$ and $\beta$), and the latter, which lacks a fully consistent probabilistic formulation.

Decoding in traditional HMM-based systems is done to incrementally find the word sequence that maximizes the practical scoring formula:
\begin{equation}\label{eq:hmm_score}
\text{Score}_{\text{HMM}}(\X,\W) = \log P(\X | \W) + \alpha \log P(\W) + \beta |\W|~,
\end{equation}
where $\alpha$ scales the influence of the language model, $\beta$ is the \emph{word insertion penalty} that penalizes (or rewards) the number of words to avoid overgeneration or undergeneration, and $|\W|$ is the number of words in the hypothesis. This formulation allows a beam search or Viterbi decoder to efficiently explore the space of possible word sequences (as expressed by the HMM states) and choose the one with the highest combined score \cite{rabiner1993fundamentals}. 

In contrast to HMMs, acoustic models that are based on wav2vec 2.0 \cite{baevski2020wav2vec} and HuBERT \cite{hsu2021hubert} are trained using the CTC loss function \cite{graves2006connectionist} on characters. During decoding of these models, a prefix beam search decoding is used, and the scoring function for each candidate word sequence $\W$ is formulated as
\begin{equation}\label{eq:ctc_score}
\text{Score}_{\text{CTC}}(\X,\W) = \log P_{\text{CTC}}(\W | \X) + \alpha \log P(\W) + \beta  |\W|~,
\end{equation}
where we assumed that the posterior probability over the characters $P(Y|X)$ can be easily translated into a probability over words $P_{\text{CTC}}(\W|\X)$. Note that systems based on these acoustic models incorporate language models by multiplying the acoustic model $P_{\text{CTC}}(\W|\X)$ by $P(\W)$  \cite{schneider2019wav2vec, baevski2020wav2vec}. This scoring function is the standard shallow-fusion formulation. While this approach yields high performance, it is implausible as it undermines the probabilistic meaning of the models.

 \section{Proposed Approach}
\label{sec:model}

We propose integrating the LLM directly into the decoding process iteratively. At each step, the LLM proposes a set of candidate tokens, and each proposal is acoustically evaluated by matching it to the speech signal by using the SSL model and combining its alignment likelihood with the LLM probability inside the MAP recursion. This ensures that only candidates with high scores from both the LLM and acoustic models remain in the beam. Unlike post-processing approaches, which operate on a completed transcript, our method incorporates the LLM within the decoding process itself. As we will explain next, this integration is implemented efficiently using dynamic programming.

We propose integrating the LLM directly into the decoding process iteratively. At each step, the LLM proposes a set of candidate tokens, and each proposal is acoustically evaluated by matching it to the speech signal by using the SSL model and combining its alignment likelihood with the LLM probability inside the MAP recursion. This ensures that only candidates with high scores from both the LLM and acoustic models remain in the beam. Unlike post-processing approaches, which operate on a completed transcript, our method incorporates the LLM within the decoding process itself. Prior work~\cite{ma2024asr} identifies tokenizer and segmentation mismatches between ASR systems and LLMs as a challenge for integrated decoding; our alignment-based scoring provides a direct way to address this issue. As we will explain next, this integration is implemented efficiently using dynamic programming.

From this point onward, we will use tokens instead of words; thus, $w_n \in \mathcal{V}$ denotes a token from the set of all tokens $\mathcal{V}$. Define the alignment sequence $A=(a_1,\ldots,a_N)$, where $a_n \in[0,T-1]$ represents the start time of token $w_n$ for $1 \le n \le N$. We denote by $\A(\W)$ the set of all possible alignments (i.e., start times) corresponding to a given token sequence $\W$. Lastly, denote by $a_1^n=(a_1,\ldots,a_n)$ the alignment (start times) of the first $n$ tokens, $w_1^n=(w_1,\ldots,w_N)$.

Our objective is to derive a MAP estimate of the most likely token sequence, $\W^*$. Rather than incorporating the probability $P(\W)$ as a whole, we decompose it into a sequence of token-level predictions. At each step, we evaluate candidate tokens by combining their LLM likelihoods with an alignment-based acoustic score derived from the SSL emissions. More formally, we start from the MAP estimator given in \eqref{eq:map1}:
\begin{eqnarray}\label{eq:map2}
\W^* &=& \arg\max_{\W} ~ P(\W | \X) \\
&=& \arg\max_{\W} \!\!\sum_{A\in\A(\W)} \!\! P(\W, A|\X).
\end{eqnarray}
We approximate this expression by assuming that the most likely alignment dominates the total probability. That is, instead of summing over all valid alignments, we approximate this sum by the alignment with the highest posterior probability,
\begin{equation}\label{eq:map_with_a}
\W^* \simeq \arg \max_{\W} \max_{A\in\A(\W)} P(\W, A \mid \X)~,
\end{equation}
following the same approximation used in the original CTC decoding formulation \cite{pmlr-v32-graves14}. 

This probability can be decomposed into two terms: a term that is associated with the acoustic model and a term that is related to the language model. We start by expressing the sequences $\W$ and $A$ explicitly: 
\begin{align}\nonumber
     \W^* &= \arg \max_{\W, A} \log P(\W, A|\X) \\ \nonumber
     &= \arg \max_{\W, A}  \sum_{n=1}^{N} \log P(w_n, a_n | w_1^{n-1}, a_1^{n-1} , \X) ~.
\end{align}
Using conditional probability, we have
\begin{align}\nonumber
     \W^* &= \arg \max_{\W, A}  \sum_{n=1}^{N} \log P(a_n | w_n, a_1^{n-1} , w_1^{n-1}, \X) \\ \label{eq:map_decomp}
     & ~~~~~~~+ \sum_{n=1}^{N} \log P(w_n | w_1^{n-1}, a_1^{n-1}, \X) ~.
\end{align}
The first term, $P(a_n | w_n, a_1^{n-1} , w_1^{n-1}, \X)$, is the probability of the start-time of a candidate token $w_n$ with respect to the speech signal $\X$ given the previous start-times $a_1^{n-1}$, and the current $w_n$ and previous tokens $ w_1^{n-1}$. This probability allows us to select the most probable token candidate and match it to the emission probabilities. We will soon describe how this probability is computed practically.

The second term, $P(w_n | w_1^{n-1}, a_1^{n-1}, \X)$, is the probability of the LLM predicting the next token, given the previous tokens and their alignments. One might consider the sampling from the LLM independent of the start-times $a_1^{n-1}$ and the input speech $\X$. However, we implicitly included them here, as the acoustic probability plays a significant role in maximizing and selecting the predicted tokens, and the previous tokens are chosen based on their alignment score.

In summary, we reformulated the MAP estimator as the sum of two terms corresponding to the acoustic and language models. The acoustic model term is expressed as $\sum_n P(a_n | w_n, a_1^{n-1} , w_1^{n-1}, \X)$, in contrast to the conventional forms $P(\X|\W)$ or $P_{\text{CTC}}(\W | \X)$. The language model term is given by $\sum_n P(w_n | w_1^{n-1}, a_1^{n-1}, \X)$, as opposed to the standard formulation $P(\W)=\sum_n P(w_n | w_1^{n-1})$. It is important to note that these terms inherently assume access to the entire speech signal $\X$, which limits the applicability of the proposed method in streaming scenarios. A simple workaround is to process speech in asynchronous chunks, allowing for partial processing. While more sophisticated strategies could be developed to fully enable streaming functionality, such approaches are beyond the scope of this work.

\section{Algorithmic Implementation}

We proceed to show that the MAP estimate in \eqref{eq:map_decomp} can be evaluated iteratively, allowing the probabilities to be computed incrementally at each step.
Equation~\eqref{eq:map_recursion} defines the incremental score of a
single partial hypothesis. In practice, the decoder applies this update
to all hypotheses maintained in a beam of size $B$, and keeps only the
highest-scoring extensions at each step (see Algorithm~\ref{alg:llm_decoder}).

Suppose we have already computed the probability up to token $n-1$, that is, we know $\log P(w_1^{n-1}, a_1^{n-1}|\X)$. The probability of token $n$ is obtained by incrementally adding two terms: the language model probability, $\log  P(w_n | w_1^{n-1}, a_1^{n-1}, \X)$ and the acoustic alignment probability  $\log  P(a_n | w_n, a_1^{n-1}, w_1^{n-1}, \X)$. Overall, we have,
\begin{align}\nonumber
\log P(w_1^n, a_1^n|\X) &=  \log P(a_n, a_1^{n-1}, w_1^n | \X)\\ \nonumber
&=\log P(a_n | a_1^{n-1}, w_1^{n}, \X) \\\nonumber
&~~~~~~~~ + \log  P(a_1^{n-1}, w_1^{n} | \X)\\ \nonumber
&=\log  P(a_n | w_n, a_1^{n-1}, w_1^{n-1}, \X) \\  \nonumber
&~~~~~~~~+ \log  P(w_n | w_1^{n-1}, a_1^{n-1}, \X) \\ \label{eq:map_recursion}
&~~~~~~~~~~~~~+ \log P(w_1^{n-1}, a_1^{n-1}|\X) . 
\end{align}

The iterative procedure operates as follows. At step $n$, we prompt the LLM with the sequence $w_1^{n-1}$ (already aligned to the speech) and sample $K$ candidate tokens $\{w_n^{(k)}\}_{k=1}^{K}$, each with an associated probability $P(w_n^{(k)} | w_1^{n-1}, a_1^{n-1}, \X)$. Each candidate token is then aligned to the speech signal $\X$, assuming the last token began at $a_{n-1}$. The alignment is performed using the Viterbi algorithm \cite{kurzinger2020ctc}, which computes the most likely alignment of each token's characters to the CTC emissions\footnote{For example, this can be implemented by performing forced alignment of the tokens $w_n$ to the CTC emission outputs of wav2vec 2.0 or HuBERT.}. For each candidate, we compute the combined probability of the token and its alignment, namely $P(w_n | w_1^{n-1}, a_1^{n-1}, \X)$ and $P(a_n | w_n, a_1^{n-1}, w_1^{n-1}, \X)$, which together define the
incremental score in Equation~\eqref{eq:map_recursion} and Equation~\eqref{eq:llm_score}. In implementation, this update is applied to all partial hypotheses in the beam, and the top $B$ extensions are retained (see Algorithm~\ref{alg:llm_decoder}).

As is standard practice in ASR decoding, a modified version of the MAP estimator is employed, as shown in \eqref{eq:hmm_score} for HMM-based decoding and \eqref{eq:ctc_score} for CTC-based decoding. Following this convention, our proposed method also introduces a scaling factor $\alpha$ for the language model probability, along with a \emph{token insertion bonus} $\beta$, analogous to the word insertion bonus commonly used in ASR. The specific choices of $\alpha$ and $\beta$ are discussed in Section~\ref{subsec:hyperparameters}. The resulting adjusted MAP objective takes the form:
\begin{align} \nonumber
     &\text{Score}_{\text{LLM}}(\X,\W) = ~ \sum_{n=1}^{N} \log P(a_n | w_n, a_1^{n-1} , w_1^{n-1}, \X) \\ \label{eq:llm_score}
     & ~~~~ + \alpha \, \sum_{n=1}^{N} \log P(w_n | w_1^{n-1}, a_1^{n-1}, \X)  +\beta \, L(w_1^{N})~.
\end{align}

\begin{algorithm}[!t]
\caption{Zero-Shot LLM-Driven ASR Decoder}
\label{alg:llm_decoder}
\begin{algorithmic}[1]
\STATE \textbf{Input:} speech signal $\mathbf{X}$ (length $T$ frames); acoustic model $\mathrm{AM}$; language model $\mathrm{LLM}$; beam width $B$; number of candidates $K$; weight $\alpha$; bonus $\beta$
\STATE Initialize $\mathcal{B}_0 \leftarrow \{(\epsilon, \epsilon, 0)\}$ \COMMENT{tokens, alignments, score}
\FOR{$n = 1$ to $N_{\max}$}
    \STATE $\mathcal{C} \leftarrow \emptyset$
    \FORALL{$(w_{1}^{n-1}, a_{1}^{n-1}, s) \in \mathcal{B}_{n-1}$}
        \STATE $\{(w_n^{(k)}, p_{\mathrm{LM}}^{(k)})\}_{k=1}^K \leftarrow \mathrm{LLM.TopK}(w_{1}^{n-1}, K)$
        \FOR{$k = 1$ to $K$}
            \STATE $(\hat{a}_n, p_{\mathrm{AM}}) \leftarrow \textsc{AM.AlignToken}(w_n^{(k)}, a_{1}^{n-1}, \mathbf{X})$
            \STATE $s' \leftarrow s + \log p_{\mathrm{AM}} + \alpha \log p_{\mathrm{LM}}^{(k)} + \beta$
            \STATE $\mathcal{C} \leftarrow \mathcal{C} \cup \{(w_{1}^{n-1} + w_n^{(k)}, a_{1}^{n-1} + \hat{a}_n, s')\}$
        \ENDFOR
    \ENDFOR
    \STATE $\mathcal{B}_n \leftarrow \textsc{BEAM.TopB}(\mathcal{C}, B)$
    \IF{$\textsc{BEAM.AllEOS}(\mathcal{B}_n)$}
        \STATE \textbf{break}
    \ENDIF
\ENDFOR
\STATE \textbf{Return:} $\textsc{BEAM.TopB}(\mathcal{B}_n, 1)$ \COMMENT{best beam defines $\mathbf{W}^*$}
\end{algorithmic}
\end{algorithm}

We conclude this section by summarizing the proposed method in the form of pseudo-code, presented in Algorithm~\ref{alg:llm_decoder}. The algorithm takes as input a speech signal $\X$, an acoustic model (wav2vec 2.0 \cite{baevski2020wav2vec} or HuBERT \cite{hsu2021hubert}) denoted as \textsc{AM}, and a large language model, (e.g., LLaMA 2, Falcon or GPT-2), denoted as \textsc{LLM}. Additional input parameters include the beam size $B$, the number of candidate tokens sampled at each iteration from the LLM $K$, the LLM scaling factor $\alpha$, and the token insertion bonus $\beta$.

The algorithm maintains two main data structures. The beam at iteration $n$, denoted by $\mathcal{B}_n$ consists of a list of tuples, where each tuple contains the token sequence of a beam hypothesis, its corresponding alignment with the speech signal, and the associated score. The second structure is the candidate list $\mathcal{C}$, which has a similar format: a list of tuples representing alternative hypotheses under consideration for the next decoding step.

At each iteration, we begin by clearing the candidate list (line 7). For each tuple in the beam from the previous iteration (line 8), we prompt the LLM with the current token sequence and sample $K$ candidate tokens (line 9). For each candidate, we determine its start time using an alignment algorithm (line 11), compute the incremental score as defined in \eqref{eq:llm_score} (line 12), and add the resulting tuple to the candidate list (line 13). The main loop terminates either when all beam hypotheses at iteration $n$ have reached a stop criterion (see Section~\ref{sec:experiments}) or when a predefined safety horizon is reached. The final output is the hypothesis in the beam with the highest score. It is worth noting that identical beam prefixes are efficiently computed only once to avoid redundant processing.

The following functions are used in Algorithm~\ref{alg:llm_decoder}.
\begin{itemize}
    \setlength{\itemsep}{0pt}
    \setlength{\parskip}{0pt}
    \setlength{\parsep}{0pt}
  \item \textsc{LM.TopK}$(w_{1}^{\,n-1}, K)$ returns the $K$ most probable next tokens and their probabilities under the LLM, given a token sequence $w_{1}^{\,n-1}$.
  \item \textsc{AM.AlignToken}$(w,a_{1}^{n-1},\X)$ uses Viterbi to find the best start time $\hat a_n$ and returns the corresponding acoustic likelihood $p_{\mathrm{AM}}$.
  \item \textsc{BEAM.TopB}$(\mathcal{C}, B)$ keeps the $B$ highest-scoring hypotheses from $\mathcal{C}$.
  \item \textsc{BEAM.AllEOS} checks whether every beam reached a stop criterion (see Section~\ref{sec:experiments} for details).
\end{itemize}

\smallskip

Before presenting the empirical evaluation, we note that the proposed approach has higher computational complexity compared to standard ASR inference using an N-gram model. Specifically, the computation of our method is of the order $O(T^2 K B)$, where $T$ represents the number of audio frames in the input, $K$ denotes the number of LLM token candidates, and $B$ is the beam size.

\section{Datasets}\label{sec:datasets}
Our system is evaluated on three widely recognized ASR datasets, leveraging pre-trained models that require no additional training or data collection. The evaluation datasets comprise the Wall Street Journal (WSJ0) test subset, TED-LIUM Release 3 test set, and a set of native American English speakers from the ALLSSTAR corpus \cite{Bradlow_ALLSSTAR}. 

The WSJ0 dataset comprises high-quality recordings of read speech from articles in the Wall Street Journal. It features 123 speakers and a balanced gender representation, making it a standard benchmark for structured speech recognition. 

The TED-LIUM Release 3 test set consists of TED talk recordings that capture real-world, professional speech delivery in diverse scenarios. This combination of datasets enables us to comprehensively assess our system's performance across various speaking styles while maintaining the original training configurations of the models. 

From the ALLSSTAR corpus \cite{Bradlow_ALLSSTAR}, we select recordings of 26 native English speakers, comprising a total of 3,060 utterances, to establish a controlled evaluation baseline aimed at consistent speech patterns. 

Our method performs zero-shot inference. Therefore, it does not require a training set. For WSJ0 and TED-LIUM 3, we used the standard validation and test set. For ALLSSTAR, we used a separate set of 3 speakers, each with 100 utterances, for validation, ensuring no overlap with the test set, which consisted of 23 speakers with a total of 2,760 utterances.

\section{Experiments}
\label{sec:experiments}

Our experimental framework evaluates multiple acoustic and language models in a zero-shot setting, utilizing pre-trained models without any additional fine-tuning or adaptation. We utilized two state-of-the-art pre-trained acoustic models. The first is wav2vec 2.0 \cite{baevski2020wav2vec}. We evaluated two model sizes: Base (12 transformer encoder blocks with 95M parameters) and Large (24 blocks, 317M parameters). These models were pre-trained on a substantial amount of unsupervised data and subsequently fine-tuned on various supervised corpus sizes: 10 minutes, 100 hours, and 960 hours. The second acoustic model is HuBERT \cite{hsu2021hubert}, which comes in two model sizes: Large (24 encoder blocks, 300M parameters) and X-Large (48 encoder blocks, 1B parameters). All acoustic models were used in their off-the-shelf configurations.

We investigated three LLMs as language models: GPT-2 \cite{radford2019language}, Falcon \cite{penedo2023refinedweb}, and LLaMA 2 \cite{touvron2023llama}. Note that each LLM has a different set of tokens, all of which are BPE-based \cite{sennrich-etal-2016-neural}. Hence, to match the different token sets to the acoustic models, we filtered each set to include only tokens comprising English alphabetic characters, excluding numeric and special characters, to ensure compatibility with the acoustic model character set. 

The pre-processing pipeline includes adding a 0.5-second silence period to each audio input, which extends the input context and empirically improves model generation stability. We then employed Silero Voice Activity Detector (VAD) version 5 \cite{SileroVAD} to trim irrelevant silence periods, optimizing the input for transcription. The parameters of Silero VAD included a start extension of 0.2 seconds and no end extension.

The decoding process utilizes beam search with beam size $b=5$ and considers the top $k=5,000$ candidates from the LLM's predictions. We will discuss how these parameters were empirically determined in Section \ref{sec:beam_and_k}. 
To ensure compatibility between the acoustic emissions and the language model, we restrict candidate tokens to those that fall within the acoustic model's character set. The decoder process employs three distinct stopping criteria: (i) completion assessment, which evaluates the likelihood of a hypothesis to be a complete sentence against the probability of adding any another token; (ii) validation of acoustic probability threshold, which terminates paths where new token additions result in probabilities falling below a threshold of 0.3, excluding whitespace tokens; and (iii) audio length alignment, which stops decoding when the generated transcript aligns with the entire duration of the input audio. These criteria work together to ensure both transcription completeness and acoustic fidelity.

\begin{table}
\centering
\caption{Model Hyper-parameters across datasets and LLMs. The parameter $\alpha$ represents the acoustic model weight, and $\beta$ is the length reward in the decoding process. These parameters remain consistent across all acoustic models for each language model-dataset pair due to the shared CTC loss training objective.}
\label{tab:model_params}
\begin{tabular}{llcc}
\toprule
Dataset & LM & $\alpha$ & $\beta$ \\
\midrule
\multirow{3}{*}{WSJ0} & LLaMA 2 & 0.0650 & 0.0051 \\
 & Falcon & 0.0949 & 0.0073 \\
 & GPT-2 & 0.0626 & 0.0090 \\
\midrule
\multirow{3}{*}{TED-LIUM 3} & LLaMA 2 & 0.0679 & 0.0028 \\
 & Falcon & 0.0695 & 0.0015 \\
 & GPT-2 & 0.0689 & 0.0061 \\
\midrule
\multirow{3}{*}{ALLSSTAR Eng} & LLaMA 2 & 0.0694 & 0.0331 \\
 & Falcon & 0.1379 & 0.0161 \\
 & GPT-2 & 0.0999 & 0.0449 \\
\bottomrule
\end{tabular}
\end{table}

Our implementation employed text normalization to ensure consistent evaluation. All texts were converted to lowercase, retaining only English alphabet characters. When processing acronyms, we merged consecutive single letters separated by spaces into a unified representation. We ensured a fair comparison between our system and beam search outputs by converting all acronyms to a standardized lowercase merged format, while addressing format variations, such as the dots in WSJ0's reference texts.

\subsection{Hyperparameters}\label{subsec:hyperparameters}

The hyperparameters $\alpha$ and $\beta$ were tuned on the validation set for each dataset and for each LLM. The values of $\alpha$ and $\beta$ are shown in Table \ref{tab:model_params}. The values of these hyperparameters remain consistent across different acoustic models. The consistency stems from the shared CTC loss function used in training the acoustic models, which results in similar emission probability distributions. This architectural advantage enables maintaining the same scaling parameters across different acoustic models, while only requiring adjustments for different language models with distinct vocabulary distributions, output characteristics, and various datasets.

\begin{table*}[htbp]
\centering
\caption{ASR performance comparison across different acoustic and language models on ALLSSTAR Eng, WSJ0, and TED-LIUM 3 datasets. WER and CER values indicate lower is better (\,$\downarrow$). \textbf{Bold} values are the best within an acoustic-model group; values marked with '\bestoverall{}' are the overall best for a given metric. \textit{LLM beam} rows correspond to our proposed method. All LM-based beam-search baselines use the standard shallow-fusion formulation.}

\label{tab:asr_comparison}
\resizebox{\textwidth}{!}{%
\begin{tabular}{>{\centering\arraybackslash}m{4cm} >{\centering\arraybackslash}m{3cm} >{\centering\arraybackslash}m{3cm} ccc ccc ccc}
\toprule
\multicolumn{3}{c}{} & \multicolumn{2}{c}{\textbf{ALLSSTAR Eng}} & \multicolumn{2}{c}{\textbf{WSJ0}} & \multicolumn{2}{c}{\textbf{TED-LIUM 3}} \\
\cmidrule(r){4-5}\cmidrule(r){6-7}\cmidrule(r){8-9}
\textbf{Acoustic Model} & \textbf{Decoder Type} & \textbf{Language Model} & WER $\downarrow$ & CER $\downarrow$ & WER $\downarrow$ & CER $\downarrow$ & WER $\downarrow$ & CER $\downarrow$ \\
\midrule
\multirow{7}{*}{\amfontsize{wav2vec 2.0 Base 10m}} & greedy             & --          & 57.03 & 18.93 & 58.15 & 16.67 & 60.72 & 20.97 \\
 & beam                & 4-gram      & 23.61 & 11.86 & 24.81 & 9.28 & 29.72 & 14.19 \\
 & beam                & GCNN        & 20.62 & 9.70  & 22.40 & 8.02 & 26.83 & \bestinblock{12.34} \\
 & beam                & Transformer & \bestinblock{18.49} & 11.41 & \bestinblock{17.01} & \bestinblock{7.49} & \bestinblock{23.04} & 12.80 \\
\cmidrule(lr){2-9}
 & \multirow{3}{*}{LLM beam} & Falcon  & 44.55 & 16.05 & 27.43 & 8.12 & 48.08 & 17.31 \\
 &                                  & GPT-2   & 21.75 & \bestinblock{9.52} & 26.93 & 8.77 & 33.49 & 13.81 \\
 &                                  & LLaMA 2 & 29.52 & 11.60 & 32.57 & 9.64 & 41.39 & 15.53 \\
\midrule
\multirow{7}{*}{\amfontsize{wav2vec 2.0 Large 10m}} & greedy             & --          & 58.95 & 20.76 & 60.32 & 18.22 & 62.44 & 23.21 \\
 & beam                & 4-gram      & 24.64 & 12.53 & 26.29 & 10.14 & 30.11 & 16.53 \\
 & beam                & GCNN        & 20.91 & \bestinblock{10.02} & 22.51 & 8.44 & 27.14 & \bestinblock{12.53} \\
 & beam                & Transformer & \bestinblock{18.01} & 10.11 & \bestinblock{16.35} & \bestinblock{7.46} & \bestinblock{23.73} & 13.37 \\
\cmidrule(lr){2-9}
 & \multirow{3}{*}{LLM beam} & Falcon  & 44.43 & 16.29 & 27.24 & 8.47 & 43.63 & 15.08 \\
 &                                  & GPT-2   & 22.04 & 10.26 & 24.05 & 8.61 & 30.65 & 13.21 \\
 &                                  & LLaMA 2 & 29.90 & 11.94 & 32.94 & 10.00 & 41.22 & 15.64 \\
\midrule
\multirow{7}{*}{\amfontsize{wav2vec 2.0 Base 100h}} & greedy             & --          & 11.71 & 4.50 & 13.39 & 3.16 & 18.31 & 6.80 \\
 & beam                & 4-gram      & 8.21  & 4.04 & 9.95 & 2.70 & 13.68 & 6.60 \\
 & beam                & GCNN        & 7.21  & \bestinblock{3.33} & 9.14 & 2.42 & 12.95 & 6.04 \\
 & beam                & Transformer & \bestinblock{6.49} & 3.64 & 9.58 & 3.57 & 14.51 & 8.43 \\
\cmidrule(lr){2-9}
 & \multirow{3}{*}{LLM beam} & Falcon  & 9.57 & 4.00 & 8.43 & 2.11 & 14.00 & 5.75 \\
 &                                  & GPT-2   & 8.07 & 3.63 & 8.57 & 2.20 & 12.95 & 5.77 \\
 &                                  & LLaMA 2 & 8.53 & 3.73 & \bestinblock{7.94} & \bestinblock{1.95} & \bestinblock{12.87} & \bestinblock{5.50} \\
\midrule
\multirow{7}{*}{\amfontsize{wav2vec 2.0 Large 100h}} & greedy             & --          & 12.10 & 4.75 & 13.58 & 3.25 & 18.82 & 7.00 \\
 & beam                & 4-gram      & 8.64 & 4.23 & 10.10 & 2.74 & 13.90 & 6.69 \\
 & beam                & GCNN        & 7.49 & \bestinblock{3.41} & 9.29 & 2.46 & 13.07 & 6.19 \\
 & beam                & Transformer & \bestinblock{6.54} & 3.72 & 9.67 & 3.51 & 14.63 & 8.57 \\
\cmidrule(lr){2-9}
 & \multirow{3}{*}{LLM beam} & Falcon  & 9.29 & 4.12 & 8.74 & 2.32 & 14.15 & 5.93 \\
 &                                  & GPT-2   & 8.16 & 3.72 & 8.89 & 2.33 & 13.05 & 6.04 \\
 &                                  & LLaMA 2 & 8.53 & 3.85 & \bestinblock{8.06} & \bestinblock{1.97} & \bestinblock{12.74} & \bestinblock{5.61} \\
\midrule
\multirow{7}{*}{\amfontsize{wav2vec 2.0 Base 960h}} & greedy             & --          & 8.25 & 3.49 & 8.86 & 1.84 & 13.68 & 5.31 \\
 & beam                & 4-gram      & 6.66 & 3.32 & 7.80 & 1.86 & 11.62 & 5.43 \\
 & beam                & GCNN        & 5.42 & 2.45 & 6.85 & 1.56 & 11.06 & 4.95 \\
 & beam                & Transformer & \bestinblock{4.64} & \bestinblock{2.44} & 8.23 & 2.41 & 12.39 & 6.75 \\
\cmidrule(lr){2-9}
 & \multirow{3}{*}{LLM beam} & Falcon  & 8.06 & 3.20 & \bestinblock{5.70} & \bestinblock{1.19} & 10.71 & 4.57 \\
 &                                  & GPT-2   & 6.24 & 2.85 & 6.46 & 1.46 & 11.15 & 4.87 \\
 &                                  & LLaMA 2 & 6.49 & 2.98 & 5.96 & 1.21 & \bestinblock{10.37} & \bestinblock{4.46} \\
\midrule
\multirow{7}{*}{\amfontsize{wav2vec 2.0 Large 960h}} & greedy             & --          & 8.48 & 3.63 & 8.99 & 1.86 & 13.84 & 5.45 \\
 & beam                & 4-gram      & 6.83 & 3.37 & 7.92 & 1.89 & 11.73 & 5.53 \\
 & beam                & GCNN        & 5.60 & 2.50 & 6.95 & 1.59 & 11.19 & 5.02 \\
 & beam                & Transformer & \bestinblock{4.73} & \bestinblock{2.47} & 8.27 & 2.43 & 12.54 & 6.87 \\
\cmidrule(lr){2-9}
 & \multirow{3}{*}{LLM beam} & Falcon  & 8.20 & 3.26 & \bestinblock{5.79} & \bestinblock{1.22} & 10.84 & 4.63 \\
 &                                  & GPT-2   & 6.32 & 2.89 & 6.56 & 1.48 & 11.26 & 4.93 \\
 &                                  & LLaMA 2 & 6.57 & 3.02 & 6.05 & 1.23 & \bestinblock{10.46} & \bestinblock{4.50} \\
\midrule
\multirow{7}{*}{\amfontsize{HuBERT Large}} & greedy             & --          & 8.11 & 3.44 & 8.61 & 1.83 & 13.53 & 5.30 \\
 & beam                & 4-gram      & 6.48 & 3.11 & 7.60 & 1.86 & 11.20 & 5.43 \\
 & beam                & GCNN        & 4.02 & \bestoverall{1.60} & 6.07 & 1.26 & 8.05 & 3.36 \\
 & beam                & Transformer & \bestoverall{3.47} & 1.64 & 7.92 & 2.35 & 9.61 & 5.01 \\
\cmidrule(lr){2-9}
 & \multirow{3}{*}{LLM beam} & Falcon  & 8.74 & 3.40 & 6.18 & 1.28 & 9.04 & 3.59 \\
 &                                  & GPT-2   & 5.36 & 2.18 & 5.53 & 1.12 & 8.13 & 3.22 \\
 &                                  & LLaMA 2 & 4.67 & 2.26 & 4.11 & \bestinblock{0.75} & \bestinblock{6.87} & \bestinblock{2.73} \\
\midrule
\multirow{7}{*}{\amfontsize{HuBERT X-Large}} & greedy             & --          & 7.96 & 3.39 & 8.48 & 1.80 & 13.27 & 5.24 \\
 & beam                & 4-gram      & 6.39 & 3.07 & 7.48 & 1.83 & 11.03 & 5.25 \\
 & beam                & GCNN        & 4.02 & \bestinblock{1.60} & 6.07 & 1.26 & 8.05 & 3.36 \\
 & beam                & Transformer & \bestinblock{3.50} & 1.66 & 7.85 & 2.31 & 9.54 & 4.98 \\
\cmidrule(lr){2-9}
 & \multirow{3}{*}{LLM beam} & Falcon  & 8.68 & 3.38 & 6.12 & 1.27 & 8.97 & 3.59 \\
 &                                  & GPT-2   & 5.32 & 2.17 & 5.49 & 1.11 & 8.09 & 3.21 \\
 &                                  & LLaMA 2 & 4.67 & 2.26 & \bestoverall{4.04} & \bestoverall{0.72} & \bestoverall{6.81} & \bestoverall{2.73} \\
\bottomrule
\end{tabular}
}
\end{table*}

\subsection{Algorithmic Optimizations}\label{sec:algo_opt}
We implemented several optimizations to improve computational efficiency. To accelerate iterative decoding, we reuse the LLM's cached key-value $(K, V)$ states as the prompt evolves, avoiding repeated full-prompt attention computations and reducing both runtime and memory usage.

We also implemented two main optimizations for the acoustic alignment process. Firstly, we cached the end frame and alignment scores from prior steps in the beam search process to minimize redundant calculations. Instead of recalculating alignments from the beginning, each new step utilized the cached frame minus one position, combining the stored scores with new calculations to refine the results. Empirical testing demonstrated that achieving optimal alignment accuracy required a one-frame overlap.
For the second optimization, we established a forward-looking boundary. Instead of aligning each prefix text with the entire audio sequence, we restricted the search to a maximum of 1,500 milliseconds (75 frames) ahead of the current position. These optimizations significantly reduced memory usage and computation time.

\subsection{Comparing different acoustic models and LLMs}

Table \ref{tab:asr_comparison} provides comprehensive performance comparisons across all model configurations and datasets. We evaluate four decoding baselines for each acoustic model. The first baseline, \emph{greedy}, performs decoding without any language model. The second, \emph{beam 4-gram}, uses beam search with a KenLM 4-gram language model, a beam size of 1500, language model weight 2.0, and word insertion penalty -1.0. The third, \emph{Transformer LM}, utilizes a 20-layer Transformer-based language model with a beam size of 500, a language model weight of 2.0, and a word insertion penalty of -1.0. The fourth, \emph{GCNN}, employs a convolutional language model with a beam size of 80, a beam threshold of 10, a language model weight of 1.0, and a word insertion bonus of 2.0. All LM-based beam-search baselines use the standard shallow-fusion formulation, following the configurations in \cite{baevski2020wav2vec, hannun2019tds} and using the Flashlight decoder via Fairseq's interface \cite{pratap2019wav2letter,ott2019fairseq}.

Our experiments demonstrate that combining the HuBERT X-Large acoustic model with the LLaMA 2 language model achieves the best performance across most datasets. This configuration yields the lowest WER and CER on WSJ0 (WER: 4.04, CER: 0.72) and TED-LIUM 3 (WER: 6.81, CER: 2.73). For the ALLSSTAR English dataset, the HuBERT Large with GPT-2 achieves optimal performance (WER: 4.57, CER: 2.15). These results align with the individual strengths of these models: HuBERT's superior acoustic modeling capabilities in the speech domain and the advanced language understanding demonstrated by LLaMA 2 in their respective evaluations.

The impact of pre-training data volume on model performance is significant, as shown in \ref{tab:asr_comparison}. Models trained on limited data (10-minute subset) consistently underperform across all decoding strategies, with WER increases of up to 31.35 for wav2vec 2.0 Base. However, as the pre-training data volume increases to 100 hours and 960 hours, our approach demonstrates substantial improvements over the baselines, highlighting the importance of sufficient pre-training data for effectively integrating large language models.

Interestingly, GCNN and Transformer LMs outperform our LLM-guided decoder on the ALLSSTAR dataset. This is likely due to the dataset's characteristics-short, syntactically simple utterances with minimal vocabulary and limited context, which diminish the benefits of long-range reasoning. In contrast, on TED-LIUM~3, which includes longer utterances, named entities, and syntactic structure, our LLM-guided decoder consistently outperforms traditional LMs, highlighting its ability to model global context and semantic dependencies.

% 3. Single column Side by Side
\begin{figure*}[ht]
    \centering
    \begin{minipage}{0.49\textwidth}
        \centering
        \includegraphics[width=\textwidth]{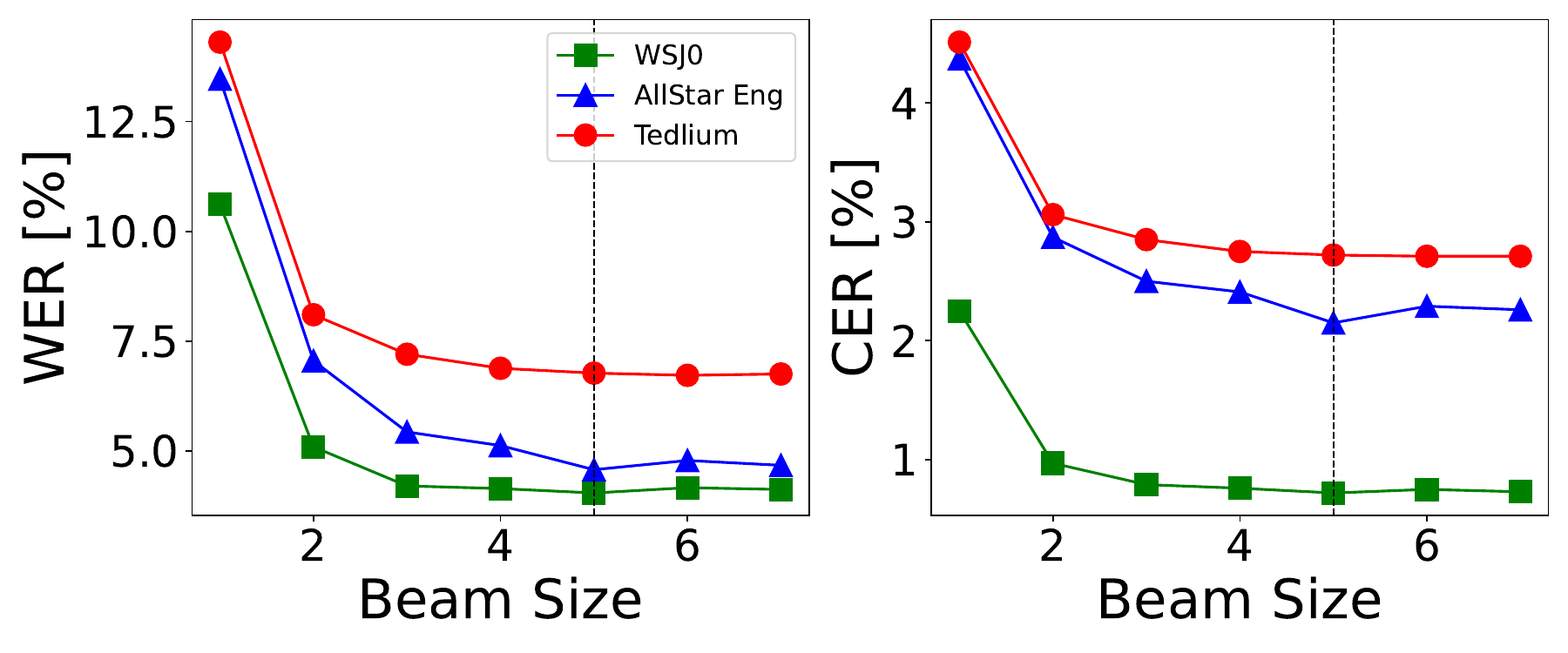}
        \caption{WER and CER using HuBERT XLarge with LLaMA 2 for different beam sizes, where beam size denotes the number of top-scoring hypotheses retained at each decoding iteration. Results are shown for WSJ0, ALLSSTAR English, and TED-LIUM datasets.}
        \label{fig:beam_size_to_wer_cer}
    \end{minipage}
    \hfill
    \begin{minipage}{0.49\textwidth}
        \centering
        \includegraphics[width=\textwidth]{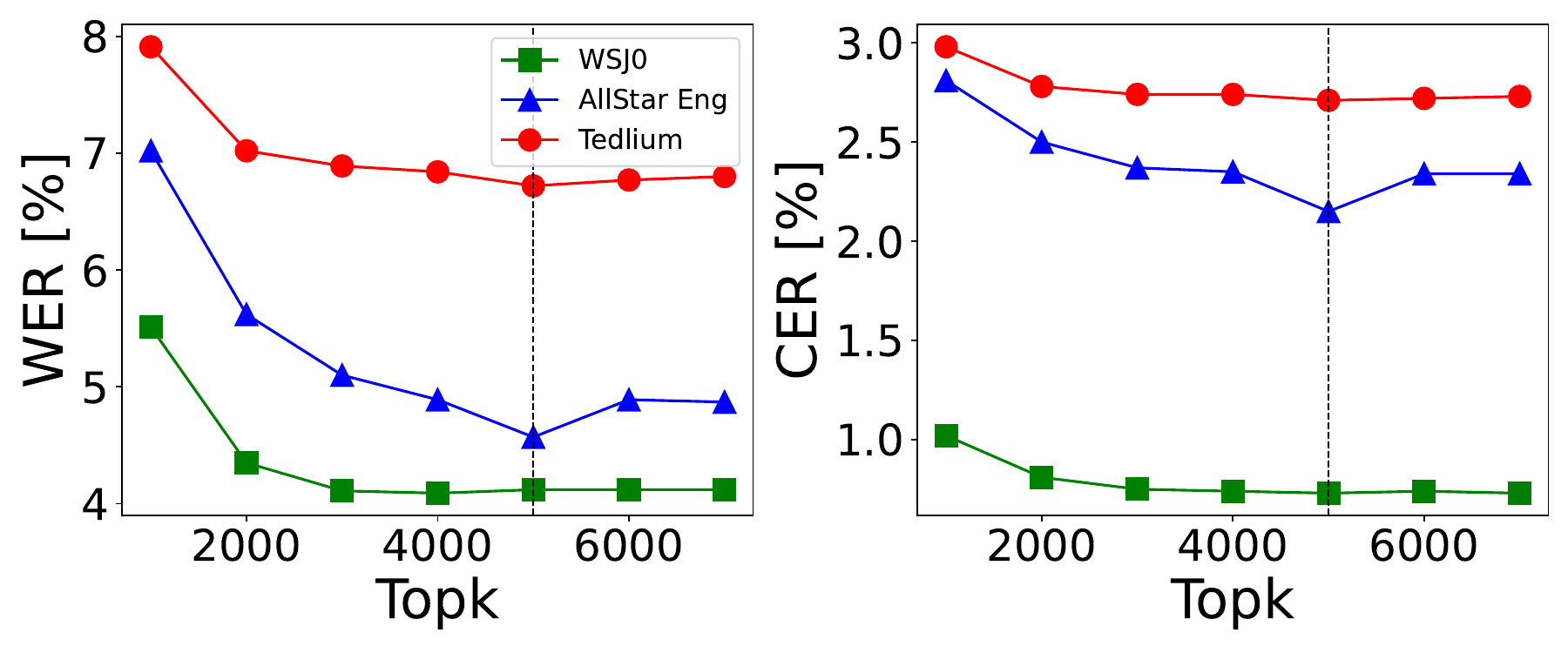}
        \caption{WER and CER using HuBERT XLarge with LLaMA 2 for different top-$k$ values, where $k$ represents the number of most probable next-token candidates the LLM considers for each prefix during transcription. Results are shown for WSJ0, ALLSSTAR English, and TED-LIUM datasets.}
        \label{fig:top_k_to_wer_cer}
    \end{minipage}
\end{figure*}

\subsection{Beam size and number of LLM candidates}\label{sec:beam_and_k}

We evaluated the influence of the LLM beam size and the number of token candidates from the LLMs. The optimal configuration parameters were determined through empirical analysis, as shown in Figures \ref{fig:beam_size_to_wer_cer} and \ref{fig:top_k_to_wer_cer}. A beam size of 5 and top-$k$ value of 5,000 were selected based on the performance plateaus observed in these analyses. The dataset-specific hyperparameters ($\alpha$, $\beta$) were tuned for each language model-dataset combination, as detailed in Table \ref{tab:model_params}, maintaining consistency across acoustic models due to their shared CTC training objective.

\subsection{Analyzing acronyms}

\begin{table*}[htbp]
\centering
\caption{Examples of acronym transcriptions by different models. (V), (X), and '-' indicate correct, incorrect, and missing predictions, respectively.}
\label{tab:acronym_examples}
\scriptsize
\begin{tabular}{m{5.2cm} m{2.0cm} m{2.0cm} m{2.0cm} m{2.0cm} m{2.0cm}}
\toprule
\textbf{Reference} & \textbf{Greedy} & \textbf{Beam 4-gram} & \textbf{GCNN} & \textbf{Transformer} & \textbf{Ours} \\ \midrule

% 1) NASA
\acronym{NASA} scheduled the launch of the space shuttle discovery for september twenty ninth.
& \textbf{NASA} (V)\,scheduled
& \textbf{A} (X)\,scheduled
& \textbf{NASA} (V)\,scheduled
& \textbf{--} (X)\,scheduled
& \textbf{NASA} (V)\,scheduled
\\ \midrule

% 2) MICC
\acronym{MICC} said it intends to pay the dividend arrears on july thirty first to stock of record july second.
& \textbf{MICC} (V)\,said
& \textbf{I} (X)\,said
& \textbf{MICC} (V)\,said
& \textbf{I} (X)\,said
& \textbf{MICC} (V)\,said
\\ \midrule

% 3) RLI twice
\acronym{RLI} corporation a peoria illinois based insurance holding company will begin trading friday on the big board under the symbol \acronym{RLI}.
& \begin{tabular}[c]{@{}l@{}} 
    \textbf{RLI} (V)\,corporation\\
    symbol \textbf{RLI} (V)
  \end{tabular}
& \begin{tabular}[c]{@{}l@{}} 
    \textbf{--} (X)\,corporation\\
    symbol \textbf{--} (X)
  \end{tabular}
& \begin{tabular}[c]{@{}l@{}} 
    \textbf{RLI} (V)\,corporation\\
    symbol \textbf{RLI} (V)
  \end{tabular}
& \begin{tabular}[c]{@{}l@{}} 
    \textbf{--} (X)\,corporation\\
    symbol \textbf{--} (X)
  \end{tabular}
& \begin{tabular}[c]{@{}l@{}} 
    \textbf{RLI} (V)\,corporation\\
    symbol \textbf{RLI} (V)
  \end{tabular}
\\ \midrule

% 4) MBA
His \acronym{MBA} also irks some colleagues who are contemptuous of foreign concepts.
& His \textbf{MBA} (V)\,also
& His \textbf{--} (X)\,also
& His \textbf{MB} (X)\,also
& His \textbf{--} (X)\,also
& His \textbf{MBA} (V)\,also
\\ \midrule

% 5) BASF + US
Two years ago \acronym{BASF} made three separate acquisitions in the \acronym{US}.
& \begin{tabular}[c]{@{}l@{}} 
    ago \textbf{BASF} (V)\,made\\
    the \textbf{US} (V)
  \end{tabular}
& \begin{tabular}[c]{@{}l@{}} 
    ago \textbf{BASF} (V)\,made\\
    the \textbf{--} (X)
  \end{tabular}
& \begin{tabular}[c]{@{}l@{}} 
    ago \textbf{BASF} (V)\,made\\
    the \textbf{US} (V)
  \end{tabular}
& \begin{tabular}[c]{@{}l@{}} 
    ago \textbf{I} (X)\,made\\
    the \textbf{US} (V)
  \end{tabular}
& \begin{tabular}[c]{@{}l@{}} 
    ago \textbf{BASF} (V)\,made\\
    the \textbf{US} (V)
  \end{tabular}
\\ \bottomrule

\end{tabular}
\end{table*}

\begin{table}[!t]
\centering
\caption{WER and CER on WSJ0 With and Without Acronym.}
\label{tab:wer_cer_results}
%\resizebox{\columnwidth}{!}{%
\begin{tabular}{lcccc}
\toprule
\multirow{2}{*}{Model} & \multicolumn{2}{c}{w/ Acronyms} & \multicolumn{2}{c}{w/o Acronyms} \\
\cmidrule(lr){2-3}\cmidrule(lr){4-5}
 & WER & CER & WER & CER \\
\midrule
Greedy      & 6.46 & 1.34 & 5.66 & 1.03 \\
4-gram      & 9.65 & 2.99 & 5.43 & 1.20 \\
GCNN        & 8.23 & 1.88 & 6.53 & 1.69 \\
Transformer & 12.74 & 5.25 & 7.58 & 2.66 \\
\textbf{Ours} & \textbf{4.78} & \textbf{0.98} & \textbf{3.86} & \textbf{0.68} \\
\bottomrule
\end{tabular}%}
\end{table}

% small space between tables

\begin{table}[!t]
\centering
\caption{Per-Acronym Recognition Accuracy on WSJ0. Accuracy is computed as the percentage of correctly transcribed acronyms by each decoding strategy.}
\label{tab:model_accuracy}
%\small  % keeps it compact but NOT stretched
\begin{tabular}{lc}
\toprule
Model       & Accuracy \\
\midrule
\textbf{Greedy}      & \textbf{0.93} \\
4-gram       & 0.39 \\
GCNN         & 0.70 \\
Transformer  & 0.27 \\
Ours         & 0.88 \\
\bottomrule
\end{tabular}
\end{table}

Acronyms, words formed from the initial letters of a phrase and pronounced as individual characters (e.g., "U-S-A"), present a unique challenge for ASR systems. In WSJ0, acronyms are formatted with dots between letters (e.g., "u. s. a."), though these dots are not valid symbols for the acoustic models' character set. Although acoustic models can represent the phonetic sequences of acronyms, traditional n-gram language models often struggle with them. Not only do these models rely heavily on predefined vocabularies with limited ability to handle out-of-vocabulary (OOV) words, but they also output acronyms as single lowercase words, making it impossible to distinguish them from regular words in the text.

The integration of a BPE-based LLM significantly enhances the system's ability to transcribe acronyms. Unlike n-gram models, BPE tokenization includes single characters as tokens, allowing the LLM to construct OOV acronyms from their individual phonetic components. This capability aligns well with the acoustic model's ability to represent sequences of single, pronounced characters. While greedy decoding can also identify acronyms correctly, it does so at the cost of significantly reduced overall transcription performance, as reflected in higher WER and CER values.

This advantage is evident in the results shown in Table~\ref{tab:wer_cer_results}. Our method achieves a WER of 4.78 and CER of 0.98 on sentences with acronyms, outperforming all other baselines. As shown in Table~\ref{tab:model_accuracy}, our method achieves an acronym recognition accuracy of 0.88, outperforming the 4-gram (0.39), GCNN (0.70), and Transformer (0.27) baselines, and approaching the accuracy of the greedy decoder (0.93). Representative transcription examples illustrating these trends are provided in Table~\ref{tab:acronym_examples}. Representative examples of acronym transcriptions across decoding strategies are shown in Table~\ref{tab:acronym_examples}.

\subsection{Semantic fidelity evaluation}
\label{subsec:semantic_bertscore}

To complement the word-level accuracy results, we evaluated the semantic fidelity of each decoding strategy using the BERTScore $F_{\text{BERT}}$ measure~\cite{zhang2019bertscore}, which computes token-level semantic similarity based on contextualized BERT embeddings. Table~\ref{tab:semantic_bertscore} reports results on WSJ0, TEDLIUM, and ALLSTAR\text{-}ENG, with WSJ0 further split into sentences with and without acronyms. The proposed decoder achieves the highest $F_{\text{BERT}}$ across all datasets, with the largest gains on acronym-containing sentences where conventional decoders often mis-segment or expand abbreviations.

Wilcoxon signed-rank tests confirm that these improvements are statistically significant (WSJ0: $p < 0.001$; TEDLIUM: $p < 0.001$; ALLSTAR-ENG: $p < 0.0001$). Even under matched-WER conditions ($|\Delta\text{WER}| \le 0.01$), the proposed decoder maintains positive semantic gains (e.g., WSJ0 acronym subset: $\Delta F_{\text{BERT}} = 0.008$-$0.014$, $p < 0.01$), indicating improved meaning preservation beyond what is captured by word-level metrics alone.

\begin{table*}[htbp]
\centering
\caption{Semantic fidelity comparison using BERTScore $F_{\text{BERT}}$ across WSJ0, TEDLIUM, and ALLSTAR\text{-}ENG. For WSJ0, results are additionally reported for sentences with and without acronyms. The proposed decoder achieves the highest $F_{\text{BERT}}$ on all datasets, reflecting improved semantic similarity to the reference transcripts.}

\label{tab:semantic_bertscore}
{%
\begin{tabular}{cccccc}
\toprule
 & \multicolumn{3}{c}{\textbf{WSJ0}} & \textbf{TEDLIUM} & \textbf{ALLSTAR-ENG} \\
\cmidrule(r){2-4}
\textbf{Model} & 
\textbf{w/ Acronyms} & 
\textbf{w/o Acronyms} & 
\textbf{Overall} &
 &
 \\
\midrule
Greedy        & 0.9778 & 0.9870 & 0.9862 & 0.9764 & 0.9905 \\
4-gram        & 0.9645 & 0.9874 & 0.9852 & 0.9767 & 0.9918 \\
GCNN          & 0.9755 & 0.9866 & 0.9856 & 0.9831 & 0.9935 \\
Transformer   & 0.9620 & 0.9880 & 0.9863 & 0.9832 & 0.9936 \\
\textbf{Ours} 
                 & \textbf{0.9861} 
                 & \textbf{0.9897} 
                 & \textbf{0.9892}
                 & \textbf{0.9850}
                 & \textbf{0.9949} \\
\bottomrule
\end{tabular}
}
\end{table*}

\subsection{Error-correction baselines}
\label{subsec:error_correction}

The results above focus on integrating language models directly into the decoding process. A natural question is how such decoding-time integration compares with LLM-based \emph{post-hoc} error-correction approaches that operate on completed ASR hypotheses. To examine this, we evaluate several zero-shot error-correction baselines applied to the outputs of the HuBERT X-Large ASR system and compare them with our LLM-guided decoder.

For consistency and reproducibility, all post-hoc error-correction baselines are implemented using open-weight Llama-3 models~\cite{dubey2024llama}. The instruction-tuned variant is used for hypothesis generation, and the base model is used for scoring where applicable, following the same zero-shot prompting baseline setups.

We implement the zero-shot error-correction baselines proposed by Ma et al.~\cite{ma2023can}: a \emph{zero-shot, unconstrained} setting, in which the LLM receives an ASR $N$-best list and generates a corrected transcription; a \emph{zero-shot selective rescoring} setting, where the LLM is instructed to select a single hypothesis using a constrained XML-style interface; and a \emph{one-shot, unconstrained} variant that augments the prompt with a single in-context example. Additionally, we implement the post-hoc baseline \emph{N-best Closest Decoding}~\cite{ma2024asr}, where the unconstrained LLM-generated correction is mapped to the ASR hypothesis with the minimum Levenshtein distance. We also include a standard $N$-best rescoring baseline that reorders ASR hypotheses by LLM scores and selects the top-ranked hypothesis.

In addition, we evaluate ProGRes~\cite{tur2024progres}, which uses a multi-step post-hoc pipeline with two different LLMs applied to the ASR $N$-best list. First, the acoustic model produces an $N$-best list using CTC beam search. An instruction-tuned Llama-3 model then generates an additional hypothesis conditioned on this list. Finally, ProGRes computes language pseudo-log-likelihoods for each hypothesis using a separate Llama-3 base model, recomputes post-hoc acoustic scores via an additional CTC log-likelihood pass, and selects the final output through linear fusion of these scores. Following their setup, the LLM score interpolation weight $\alpha$ is tuned on development data. The optimal value is $\alpha = 0.1$ for ALLSSTAR-Eng, whereas for WSJ0 and TED-LIUM~3 the best performance is obtained with $\alpha = 0$, meaning the LLM-based score does not affect the selected hypothesis for these datasets.

Overall, results are summarized in Table~\ref{tab:ec_comparison}. The table reports WER and CER, along with word insertion rates (Ins\%) to reflect each method's tendency to introduce content not supported by the acoustic evidence. We additionally report \emph{relative inference overhead}, which reflects the number of extra inference operations beyond the common CTC beam-search decoding step, expressed in terms of LLM hypothesis generation, LLM scoring over an ASR $N$-best list, and additional CTC forward passes. For our decoder, integrated scoring amortizes this overhead via prefix reuse and KV caching (Section~\ref{sec:algo_opt}).

While some post-hoc baselines achieve competitive performance on ALLSSTAR-Eng, which consists of short, read utterances, performance differences become more pronounced on WSJ0 and TED-LIUM~3, which represent more complex domains with longer, structured text and spontaneous speech, respectively. Post-hoc methods do not condition generation on the acoustic signal and may therefore produce semantically reasonable outputs that are not supported by the audio, commonly referred to as hallucinations. For example, the selective rescoring variant proposed by Ma et al. performs poorly, with extremely high WER and CER, because instruction-tuned LLMs do not always follow output-format constraints. This behavior is also observed for ProGRes, which, despite achieving the lowest WER and CER on the ALLSSTAR-Eng dataset, produces hallucinations such as extending "\emph{is the kangerlusig river in southwest}" to "\emph{is the kangerlusig river in southwest Greenland},” or augmenting "\emph{they waited for an hour}" with templated continuations that append spurious metadata. This trend is also reflected in higher insertion rates for post-hoc methods, as reported in Table~\ref{tab:ec_comparison}, indicating a greater tendency to introduce acoustically unsupported content.
\begin{table*}[htbp]
\centering
\caption{Error-correction performance across ALLSSTAR-Eng, WSJ0, and TED-LIUM~3.
WER, CER, word insertion rate (Ins\%), and relative inference overhead are reported.
All methods start from ASR CTC beam-search decoding.}
\label{tab:ec_comparison}
{%
\begin{tabular}{
    l l
    cc cc cc
}
\toprule
\textbf{Method}
& \begin{tabular}[c]{@{}c@{}}
\textbf{Relative inference} \\
\textbf{overhead}
\end{tabular}
& \multicolumn{2}{c}{\textbf{ALLSSTAR Eng}}
& \multicolumn{2}{c}{\textbf{WSJ0}}
& \multicolumn{2}{c}{\textbf{TED-LIUM 3}} \\
\cmidrule(r){3-4}\cmidrule(r){5-6}\cmidrule(r){7-8}
& 
& WER (Ins\%) $\downarrow$ & CER $\downarrow$
& WER (Ins\%) $\downarrow$ & CER $\downarrow$
& WER (Ins\%) $\downarrow$ & CER $\downarrow$ \\
\midrule

0-shot uncon~\cite{ma2023can}
& 1$\times$ LLM generation
& 3.41 (0.55) & 1.85
& 5.70 (0.93) & 1.56
& 9.49 (1.70) & 5.22 \\

0-shot select~\cite{ma2023can}
& 1$\times$ LLM generation
& 1445.80 (1443.48) & 1725.74
& 436.62 (429.17) & 447.37
& 338.55 (328.96) & 374.97 \\

1-shot uncon~\cite{ma2023can}
& 1$\times$ LLM generation
& 4.17 (0.69) & 2.36
& 5.11 (0.66) & 1.29
& 7.52 (1.46) & 3.68 \\

N-best Closest Decoding~\cite{ma2024asr}
& 1$\times$ LLM generation
& 3.67 (0.40) & \textbf{1.71}
& 4.73 (0.90) & 0.93
& 7.94 (1.51) & 3.01 \\

ProGRes~\cite{tur2024progres}
& \begin{tabular}[c]{@{}l@{}}
1$\times$ LLM generation + \\
~$N\times$ LLM scoring + \\
~~$N\!\times\!\!$ CTC pass
\end{tabular}
& \textbf{3.24 (0.31)} & 1.92
& 8.46 (1.46) & 1.65
& 9.86 (1.90) & 3.32 \\

N-best rescoring~\cite{chelba2012large}
& $N\times$ LLM scoring
& 4.29 (0.49) & 2.26
& 6.72 (0.61) & 3.47
& 14.91 (1.24) & 11.01 \\

Ours
& \begin{tabular}[c]{@{}l@{}}
integrated LLM scoring \\
~with KV caching \\
~~during beam search
\end{tabular}
& 4.67 (0.68) & 2.26
& \textbf{4.04 (1.01)} & \textbf{0.72}
& \textbf{6.81 (1.62)} & \textbf{2.73} \\

\bottomrule
\end{tabular}%
}
\end{table*}

To quantify this systematically, we use the hallucination identification setup of Frieske and Shi~\cite{frieske2024hallucinations}, which flags utterances with WER above 30\%, cosine similarity below 0.2, and language-model perplexity below 200. The resulting sentence-level hallucination rates are reported in Table~\ref{tab:hallucination_rates}. Our decoder maintains near-zero hallucination rates across all datasets, whereas post-hoc methods may hallucinate, with the largest effect observed for the selective interface baseline. Perplexity is computed using a Flan-T5-small language model, and semantic similarity is measured using sentence embeddings from all-MiniLM-L6-v2

\begin{table*}[t]
\centering
\caption{Sentence-level hallucination rates across ALLSSTAR-Eng, WSJ0, and TED-LIUM~3.
Hall.\% (N) reports the percentage and absolute count of hallucinated utterances.}
\label{tab:hallucination_rates}
\setlength{\tabcolsep}{4pt}
\begin{tabular}{lccc}
\toprule
\textbf{Method}
& \textbf{ALLSSTAR Eng}
& \textbf{WSJ0}
& \textbf{TED-LIUM~3} \\
& Hall.\% (N) $\downarrow$
& Hall.\% (N) $\downarrow$
& Hall.\% (N) $\downarrow$ \\
\midrule

0-shot uncon~\cite{ma2023can}
& 0.00 (0)
& 0.00 (0)
& 0.35 (4) \\

0-shot select~\cite{ma2023can}
& 13.19 (364)
& 13.57 (90)
& 22.60 (261) \\

1-shot uncon~\cite{ma2023can}
& 0.00 (0)
& 0.00 (0)
& 0.35 (4) \\

N-best Closest Decoding~\cite{ma2024asr}
& 0.00 (0)
& 0.00 (0)
& 0.35 (4) \\

N-best rescoring~\cite{chelba2012large}
& 0.40 (11)
& 2.56 (17)
& 6.58 (76) \\

ProGRes~\cite{tur2024progres}
& 0.40 (11)
& 0.00 (0)
& 0.35 (4) \\

Ours
& 0.00 (0)
& 0.00 (0)
& 0.35 (4) \\

\bottomrule
\end{tabular}
\end{table*}

\section{Conclusions}
\label{sec:conclusions}

In this work, we introduced a novel zero-shot decoding approach that positions LLMs as the primary driver of the decoding process in SSL-ASR systems, moving away from traditional approaches dominated by acoustic models. We derived this approach theoretically from the MAP estimator of tokens given the speech signal and proposed an iterative procedure to solve it efficiently.

By leveraging LLMs as language models, the approach utilizes their advanced linguistic capabilities, such as understanding context and domain-specific vocabulary, to dynamically guide word sequence adjustments. The seamless integration of pre-trained LLMs and acoustic models without retraining enables the system to handle diverse speech patterns effectively. 

In contrast to post-hoc error correction and N-best, rescoring approaches, which apply an LLM only to the final ASR output, our method integrates the LLM directly into the decoding process. This enables a step-by-step interaction between linguistic modeling and acoustic evidence, favoring hypotheses that remain aligned with the speech signal and avoiding acoustically unsupported additions.

Experiments across multiple benchmarks demonstrated improved semantic fidelity and competitive WER relative to strong baselines, validating the robustness and adaptability of the proposed approach. At a broader level, our goal is to close the performance gap between modular ASR systems with separately trained acoustic and language models and end-to-end supervised transformer-based models such as Whisper \cite{radford2023whisper}.

The proposed approach presents two significant drawbacks, which we will address in future research. Firstly, while LLMs demonstrate substantial advantages over traditional language models across various tasks, this does not correspond to a marked improvement in the WER of the proposed ASR system. We believe this issue stems from a relatively weak acoustic model. Future work will focus on integrating more robust acoustic models, such as the encoder from Whisper \cite{radford2023whisper}. This integration should enhance both acoustic and language representations and provide a means to replace the internal, relatively weak language model of Whisper.

The second direction for future work involves adapting the current method for streaming ASR, which is crucial for real-time applications. This requires rethinking the system's computational flow, particularly by modifying the acoustic model to support incremental processing. One promising avenue is to restructure the attention mechanism-typically a bottleneck in non-streaming models-so that attention matrices can be computed and updated incrementally, frame by frame. Such advancements would enable the ASR system to produce partial transcriptions on-the-fly while maintaining the benefits of deep integration with a powerful LLM.

\bibliographystyle{IEEEtran}
\bibliography{refs}

@inproceedings{hannun2019tds,
  title     = {Sequence-to-Sequence Speech Recognition with Time-Depth Separable Convolutions},
  author    = {Hannun, Awni and Case, Cody and Casper, Jared and Catanzaro, Bryan and Diamos, Gregory and others},
  booktitle = {Proceedings of Interspeech},
  year      = {2019},
  doi       = {10.21437/Interspeech.2019-2460}
}

@inproceedings{barcovschi2023comparative,
  title={A comparative analysis between Conformer-Transducer, Whisper, and wav2vec2 for improving the child speech recognition},
  author={Barcovschi, Andrei and Jain, Rishabh and Corcoran, Peter},
  booktitle={2023 International Conference on Speech Technology and Human-Computer Dialogue (SpeD)},
  pages={42--47},
  year={2023},
  organization={IEEE}
}

@article{anidjar2024whisper,
  title={Whisper turns stronger: Augmenting wav2vec 2.0 for superior asr in low-resource languages},
  author={Anidjar, Or Haim and Marbel, Revital and Yozevitch, Roi},
  journal={arXiv preprint arXiv:2501.00425},
  year={2024}
}

@article{rouditchenko2023comparison,
  title={Comparison of multilingual self-supervised and weakly-supervised speech pre-training for adaptation to unseen languages},
  author={Rouditchenko, Andrew and Khurana, Sameer and Thomas, Samuel and Feris, Rogerio and Karlinsky, Leonid and Kuehne, Hilde and Harwath, David and Kingsbury, Brian and Glass, James},
  journal={arXiv preprint arXiv:2305.12606},
  year={2023}
}

@article{adnan2025one,
  title={Which one Performs Better? Wav2Vec or Whisper? Applying both in Badini Kurdish Speech to Text (BKSTT)},
  author={Adnan, Renas and Hassani, Hossein},
  journal={arXiv preprint arXiv:2508.09957},
  year={2025}
}

@article{chen2022wavlm,
  title={Wavlm: Large-scale self-supervised pre-training for full stack speech processing},
  author={Chen, Sanyuan and Wang, Chengyi and Chen, Zhengyang and Wu, Yu and Liu, Shujie and Chen, Zhuo and Li, Jinyu and Kanda, Naoyuki and Yoshioka, Takuya and Xiao, Xiong and others},
  journal={IEEE Journal of Selected Topics in Signal Processing},
  volume={16},
  number={6},
  pages={1505--1518},
  year={2022},
  publisher={IEEE}
}

@article{zhang2019bertscore,
  title={Bertscore: Evaluating text generation with bert},
  author={Zhang, Tianyi and Kishore, Varsha and Wu, Felix and Weinberger, Kilian Q and Artzi, Yoav},
  journal={arXiv preprint arXiv:1904.09675},
  year={2019}
}

@inproceedings{ott2019fairseq,
  author    = {M.\,Ott and S.\,Edunov and A.\,Baevski and A.\,Fan and S.\,Gross and N.\,Ng and D.\,Grangier and M.\,Auli},
  title     = {fairseq: A Fast, Extensible Toolkit for Sequence Modeling},
  booktitle = {Proc.\ Conf.\ North American Chapter of the Association for Computational Linguistics: Human Language Technologies (Demo)},
  year      = {2019},
  month     = jun,
  address   = {Minneapolis, MN, USA},
  pages     = {48--53},
  doi       = {10.18653/v1/N19-4009}
}

@InProceedings{pmlr-v162-wei22d,
  title = 	 {Mitigating Neural Network Overconfidence with Logit Normalization},
  author =       {Wei, Hongxin and Xie, Renchunzi and Cheng, Hao and Feng, Lei and An, Bo and Li, Yixuan},
  booktitle = 	 {Proceedings of the 39th International Conference on Machine Learning},
  pages = 	 {23631--23644},
  year = 	 {2022},
  editor = 	 {Chaudhuri, Kamalika and Jegelka, Stefanie and Song, Le and Szepesvari, Csaba and Niu, Gang and Sabato, Sivan},
  volume = 	 {162},
  series = 	 {Proceedings of Machine Learning Research},
  month = 	 {17--23 Jul},
  publisher =    {PMLR},
  url = 	 {https://proceedings.mlr.press/v162/wei22d.html}
}

@article{kim2024automatic,
  title={Automatic recognition of second language speech-in-noise},
  author={Kim, Seung-Eun and Chernyak, Bronya R and Seleznova, Olga and Keshet, Joseph and Goldrick, Matthew and Bradlow, Ann R},
  journal={JASA Express Letters},
  volume={4},
  number={2},
  year={2024},
  publisher={AIP Publishing}
}

@book{huang2001spoken,
  title={Spoken Language Processing: A guide to theory, algorithm, and system development},
  author={Huang, Xuedong and Acero, Alex and Hon, Hsiao-Wuen and Reddy, Raj},
  year={2001},
  publisher={Prentice hall PTR}
}

@inproceedings{toshniwal2018comparison,
  title={A comparison of techniques for language model integration in encoder-decoder speech recognition},
  author={Toshniwal, Shubham and Kannan, Anjuli and Chiu, Chung-Cheng and Wu, Yonghui and Sainath, Tara N and Livescu, Karen},
  booktitle={2018 IEEE spoken language technology workshop (SLT)},
  pages={369--375},
  year={2018},
}

@InProceedings{pmlr-v32-graves14,
  title = 	 {Towards End-To-End Speech Recognition with Recurrent Neural Networks},
  author = 	 {Graves, Alex and Jaitly, Navdeep},
  booktitle = 	 {Proceedings of the 31st International Conference on Machine Learning},
  pages = 	 {1764--1772},
  year = 	 {2014},
  editor = 	 {Xing, Eric P. and Jebara, Tony},
  volume = 	 {32},
  number =       {2},
  series = 	 {Proceedings of Machine Learning Research},
  address = 	 {Bejing, China},
  month = 	 {22--24 Jun},
  publisher =    {PMLR},
}

@inproceedings{dauphin2017language,
  author       = {Yann N. Dauphin and Angela Fan and Michael Auli and David Grangier},
  title        = {Language Modeling with Gated Convolutional Networks},
  booktitle    = {Proceedings of the 34th International Conference on Machine Learning (ICML)},
  pages        = {933--941},
  year         = {2017},
  organization = {PMLR}
}

@inproceedings{baevski2018adaptive,
  author    = {Alexei Baevski and Michael Auli},
  title     = {Adaptive Input Representations for Neural Language Modeling},
  booktitle = {Proc. Int. Conf. Learn. Representations (ICLR)},
  year      = {2019},
  address   = {New Orleans, LA, USA},
}

@article{arora2025landscape,
  author       = {Siddhant Arora and Kai-Wei Chang and Chung-Ming Chien and Yifan Peng and Haibin Wu and Yossi Adi and Emmanuel Dupoux and Hung-yi Lee and Karen Livescu and Shinji Watanabe},
  title        = {On the Landscape of Spoken Language Models: A Comprehensive Survey},
  journal={arXiv:2504.08528},
  year         = {2025},
  month        = {April},
  doi          = {10.48550/arXiv.2504.08528},
  url          = {https://arxiv.org/abs/2504.08528}
}

@inproceedings{pratap2019wav2letter,
  author    = {V.\,Pratap and A.\,Hannun and Q.\,Xu and J.\,Cai and J.\,Kahn and G.\,Synnaeve and V.\,Liptchinsky and R.\,Collobert},
  title     = {wav2letter++: The Fastest Open‑Source Speech Recognition System},
  booktitle = {Proc.\ IEEE Int.\ Conf.\ Acoust., Speech, and Signal Processing (ICASSP)},
  year      = {2019},
}

@book{jm3,
author =       "Daniel Jurafsky and James H. Martin",
title =        "Speech and Language Processing: An Introduction to
             Natural Language Processing, Computational Linguistics,
             and Speech Recognition with Language Models",
year = {2025},
url = {https://web.stanford.edu/~jurafsky/slp3/},
publisher = "Online manuscript released January 12, 2025",
edition =         "3rd",
}

@book{Jelinek1998,
  author    = {Frederick Jelinek},
  title     = {Statistical Methods for Speech Recognition},
  year      = {1998},
  publisher = {The MIT Press},
}

@inproceedings{graves2006connectionist,
  title={Connectionist temporal classification: Labelling unsegmented sequence data with recurrent neural networks},
  author={Graves, Alex and Fern{\'a}ndez, Santiago and Gomez, Faustino and Schmidhuber, J{\"u}rgen},
  booktitle={Proceedings of the 23rd international conference on Machine learning},
  doi={10.1145/1143844.1143891},
  pages={369--376},
  year={2006}
}

@inproceedings{sennrich-etal-2016-neural,
    title = "Neural Machine Translation of Rare Words with Subword Units",
    author = "Sennrich, Rico  and
      Haddow, Barry  and
      Birch, Alexandra",
    booktitle = "Proceedings of the 54th Annual Meeting of the Association for Computational Linguistics (Volume 1: Long Papers)",
    month = aug,
    year = "2016",
    address = "Berlin, Germany",
    publisher = "Association for Computational Linguistics",
    url = "https://aclanthology.org/P16-1162/",
    doi = {10.18653/v1/P16-1162},
    pages = "1715--1725"
}

@book{rabiner1993fundamentals,
  title={Fundamentals of Speech Recognition},
  author={Rabiner, Lawrence},
  publisher={Prentice Hall},
  year={1993}
}

@inproceedings{schneider2019wav2vec,
  title={wav2vec: Unsupervised pre-training for speech recognition},
  author={Schneider, Steffen and Baevski, Alexei and Collobert, Ronan and Auli, Michael},
  booktitle={Proceedings of the 20th Annual Conference of the International Speech Communication Association (INTERSPEECH)},
  doi={10.21437/interspeech.2019-1873},
  year={2019},
  publisher={ISCA}
}

@inproceedings{baevski2020wav2vec,
  title={wav2vec 2.0: A framework for self-supervised learning of speech representations},
  author={Baevski, Alexei and Zhou, Yuhao and Mohamed, Abdel-rahman and Auli, Michael},
  booktitle={Advances in Neural Information Processing Systems},
  volume={33},
  doi = {10.48550/arXiv.2006.11477},
  pages={12449--12460},
  year={2020}
}

@article{hsu2021hubert,
  title={{HuBERT}: Self-supervised speech representation learning by masked prediction of hidden units},
  author={Hsu, Wei-Ning and Bolte, Benjamin and Tsai, Yao-Hung Hubert and Lakhotia, Kushal and Salakhutdinov, Ruslan and Mohamed, Abdelrahman},
  journal={IEEE/ACM transactions on audio, speech, and language processing},
  doi={10.1109/taslp.2021.3122291},
  volume={29},
  pages={3451--3460},
  year={2021},
  publisher={IEEE}
}

@article{radford2019language,
  title={Language models are unsupervised multitask learners},
  author={Radford, Alec and Wu, Jeffrey and Child, Rewon and Luan, David and Amodei, Dario and Sutskever, Ilya},
  journal={OpenAI blog},
  volume={1},
  number={8},
  pages={9},
  year={2019},
  doi={10.48550/arXiv.1905.11485}
}

@article{touvron2023llama,
  title={{LLaMA}: open and efficient foundation language models},
  author={Touvron, Hugo and Lavril, Thibaut and Izacard, Gautier and Martinet, Xavier and Lachaux, Marie-Anne and Lacroix, Timoth{\'e}e and Rozi{\`e}re, Baptiste and Goyal, Naman and Hambro, Eric and Azhar, Faisal and Rodriguez, Aurelien and Joulin, Armand and Grave, Edouard and Lample, Guillaume},
  journal={arXiv preprint arXiv:2302.13971},
  year={2023},
  doi={10.48550/arXiv.2302.13971}
}

@inproceedings{radford2023whisper,
  title={Robust speech recognition via large-scale weak supervision},
  author={Radford, Alec and Kim, Jong Wook and Xu, Tao and Brockman, Greg and McLeavey, Christine and Sutskever, Ilya},
  booktitle={International conference on machine learning},
  pages={28492--28518},
  year={2023},
  organization={PMLR},
  url={https://arxiv.org/abs/2212.04356},
  doi={10.48550/arXiv.2212.04356}
}

@article{ma2024asr,
  title={ASR error correction using large language models},
  author={Ma, Rao and Qian, Mengjie and Gales, Mark and Knill, Kate},
  journal={arXiv preprint arXiv:2409.09554},
  year={2024},
  doi={10.48550/arXiv.2409.09554}
}

@inproceedings{tur2024progres,
  title={Progres: Prompted generative rescoring on asr N-best},
  author={Tur, Ada Defne and Moumen, Adel and Ravanelli, Mirco},
  booktitle={2024 IEEE Spoken Language Technology Workshop (SLT)},
  pages={600--607},
  year={2024},
  organization={IEEE},
  doi={10.48550/arXiv.2409.00217}
}

@article{ma2023can,
  title={Can generative large language models perform asr error correction?},
  author={Ma, Rao and Qian, Mengjie and Manakul, Potsawee and Gales, Mark and Knill, Kate},
  journal={arXiv:2307.04172},
  year={2023},
  doi={10.48550/arXiv.2307.04172}
}

@article{penedo2023refinedweb,
  title={The RefinedWeb dataset for Falcon LLM: outperforming curated corpora with web data, and web data only},
  author={Penedo, Guilherme and Malartic, Quentin and Hesslow, Daniel and Cojocaru, Ruxandra and Cappelli, Alessandro and Alobeidli, Hamza and Pannier, Baptiste and Almazrouei, Ebtesam and Launay, Julien},
  journal={arXiv:2306.01116},
  year={2023},
  doi={10.48550/arXiv.2306.01116}
}

@article{deng2024transducer,
  title={{Transducer-LLaMA}: Integrating LLMs into streamable Transducer-based speech recognition},
  author={Deng, Keqi and Guo, Jinxi and Ma, Yingyi and Moritz, Niko and Woodland, Philip C and Kalinli, Ozlem and Seltzer, Mike},
  journal={arXiv:2412.16464},
  year={2024},
  doi={10.48550/arXiv.2412.16464}
}

@article{jia2024efficient,
  title={Efficient streaming LLM for speech recognition},
  author={Jia, Junteng and Keren, Gil and Zhou, Wei and Lakomkin, Egor and Zhang, Xiaohui and Wu, Chunyang and Seide, Frank and Mahadeokar, Jay and Kalinli, Ozlem},
  journal={arXiv:2410.03752},
  year={2024},
  doi={10.48550/arXiv.2410.03752}
}

@article{ma2024embarrassingly,
  title={An embarrassingly simple approach for LLM with strong ASR capacity},
  author={Ma, Ziyang and Yang, Guanrou and Yang, Yifan and Gao, Zhifu and Wang, Jiaming and Du, Zhihao and Yu, Fan and Chen, Qian and Zheng, Siqi and Zhang, Shiliang and Chen, Xie},
  journal={arXiv:2402.08846},
  year={2024},
  doi={10.48550/arXiv.2402.08846}
}

@article{mundnich2024zero,
  title={Zero-resource speech translation and recognition with LLMs},
  author={Mundnich, Karel and Niu, Xing and Mathur, Prashant and Ronanki, Srikanth and Houston, Brady and Elluru, Veera Raghavendra and Das, Nilaksh and Hou, Zejiang and Huybrechts, Goeric and Bhatia, Anshu and Garcia-Romero, Daniel and Han, Kyu J. and Kirchhoff, Katrin},
  journal={arXiv:2412.18566},
  year={2024},
  doi={10.48550/arXiv.2412.18566}
}

@inproceedings{Bridle1989,
  title = {Probabilistic Interpretation of Feedforward Classification Network Outputs, with Relationships to Statistical Pattern Recognition},
  author={Bridle, John S.},
  booktitle={Neurocomputing},
  pages={227--236},
  year={1989},
  publisher={Springer}
}

@misc{SileroVAD,
  author       = {Silero Team},
  title        = {Silero VAD: pre-trained enterprise-grade Voice Activity Detector (VAD), Number Detector and Language Classifier},
  year         = {2024},
  publisher    = {GitHub},
  journal      = {GitHub repository},
  howpublished = {\url{https://github.com/snakers4/silero-vad}},
  email        = {hello@silero.ai}
}

@inproceedings{kurzinger2020ctc,
  title={{CTC}-segmentation of large corpora for german end-to-end speech recognition},
  author={K{\"u}rzinger, Ludwig and Winkelbauer, Dominik and Li, Lujun and Watzel, Tobias and Rigoll, Gerhard},
  booktitle={International Conference on Speech and Computer (SPECOM)},
  pages={267--278},
  year={2020},
  organization={Springer},
  doi={10.1007/978-3-030-60276-5_27}
}

@misc{Bradlow_ALLSSTAR,
  author = {Bradlow, Ann R.},
  title = {{ALLSSTAR}: Archive of L1 and L2 Scripted and Spontaneous Transcripts And Recordings},
  year ={2023},
  publisher = {Northwestern University},
  url = {https://speechbox.linguistics.northwestern.edu/allsstar}
}

@article{frieske2024hallucinations,
  title={Hallucinations in neural automatic speech recognition: Identifying errors and hallucinatory models},
  author={Frieske, Rita and Shi, Bertram E.},
  journal={arXiv preprint arXiv:2401.01572},
  year={2024},
  doi={10.48550/arXiv.2401.01572}
}

@article{chelba2012large,
  title={Large scale language modeling in automatic speech recognition},
  author={Chelba, Ciprian and Bikel, Dan and Shugrina, Maria and Nguyen, Patrick and Kumar, Shankar},
  journal={arXiv preprint arXiv:1210.8440},
  year={2012}
}

@inproceedings{xu2018pruned,
  title={A pruned rnnlm lattice-rescoring algorithm for automatic speech recognition},
  author={Xu, Hainan and Chen, Tongfei and Gao, Dongji and Wang, Yiming and Li, Ke and Goel, Nagendra and Carmiel, Yishay and Povey, Daniel and Khudanpur, Sanjeev},
  booktitle={2018 IEEE international conference on acoustics, speech and signal processing (ICASSP)},
  pages={5929--5933},
  year={2018},
  organization={IEEE}
}

@inproceedings{xu2022rescorebert,
  title={Rescorebert: Discriminative speech recognition rescoring with bert},
  author={Xu, Liyan and Gu, Yile and Kolehmainen, Jari and Khan, Haidar and Gandhe, Ankur and Rastrow, Ariya and Stolcke, Andreas and Bulyko, Ivan},
  booktitle={ICASSP 2022-2022 IEEE International Conference on Acoustics, Speech and Signal Processing (ICASSP)},
  pages={6117--6121},
  year={2022},
  organization={IEEE}
}

@article{udagawa2022effect,
  title={Effect and analysis of large-scale language model rescoring on competitive asr systems},
  author={Udagawa, Takuma and Suzuki, Masayuki and Kurata, Gakuto and Itoh, Nobuyasu and Saon, George},
  journal={arXiv preprint arXiv:2204.00212},
  year={2022}
}

@article{dubey2024llama,
  title={The llama 3 herd of models},
  author={Dubey, Abhimanyu and Jauhri, Abhinav and Pandey, Abhinav and Kadian, Abhishek and Al-Dahle, Ahmad and Letman, Aiesha and Mathur, Akhil and Schelten, Alan and Yang, Amy and Fan, Angela and others},
  journal={arXiv e-prints},
  pages={arXiv--2407},
  year={2024}
}
\end{document}